\documentclass[onecolumn,12pt]{IEEEtran}
\usepackage[margin=1in]{geometry}
\usepackage{amsmath}
\usepackage{amssymb}
\usepackage{graphicx}
\usepackage{subfig}
\usepackage{algorithm}
\usepackage[noend]{algpseudocode}
\usepackage{url}
\usepackage{hyperref}

\title{Unsupervised Machine Learning of Open Source Russian Twitter Data Reveals Global Scope and Operational Characteristics}

\author{Christopher Griffin and Brady Bickel\thanks{C. Griffin and B. Bickel are with the Applied Research Laboratory, The Pennsylvania State University, University Park, PA 16802. E-mail: griffinch@ieee.org, brb162@arl.psu.edu.}
}

\date{\today}

\begin{document}

\maketitle

\begin{abstract}
  We developed and used a collection of statistical methods (unsupervised machine learning) to extract relevant information from a Twitter supplied data set consisting of alleged Russian trolls who (allegedly) attempted to influence the 2016 US Presidential election. These unsupervised statistical methods allow fast identification of (i) emergent language communities within the troll population, (ii) the transnational scope of the operation and (iii) operational characteristics of trolls that can be used for future identification. Using natural language processing, manifold learning and Fourier analysis, we identify an operation that includes not only the 2016 US election, but also the French National and both local and national German elections. We show the resulting troll population is composed of users with common, but clearly customized, behavioral characteristics.
\end{abstract}

\section{Significance Statement}
The advent of large-scale social media and Web 2.0 technologies promised a world more connected than ever. This technology has been perverted to create systems perfect for large-scale manipulation of populations with the potential to influence the democratic process itself. In this work, we use natural language processing, manifold learning and Fourier analysis to show not only the extent of one such (alleged operation) using open source Twitter data, but also show that these influence operations have identifiable operating characteristics that (potentially) can be used for future identification.

\section*{Introduction}
The use of Web 2.0 technologies has exploded over the past 10 years \cite{J15}, with most individuals using one or more social networks \cite{P15}. The potential to communicate with family, friends and like-minded individuals is an obvious boon of these services. Unfortunately, these services have exploited the darker aspects of human psychology \cite{BWJT17} to encourage their use and have created platforms in which outside actors may attempt to influence large portions of the population. Like spam, even if a small portion of the population is influenced \cite{RR12}, it may have an outsized effect on the democratic process, threatening democracy itself.

In this paper, we analyze a data set purported to represent Russian troll twitter messages, provided by Twitter itself and obtained from NBCNews \cite{P18}. We acknowledge the broadly held assertions that Russian operatives attempted to use messaging in the social media environment to influence the 2016 US Presidential election. We make no claim to \textit{prove or refute} these assertions.  This paper is primarily exploratory in nature and provides a case study for the application of unsupervised machine learning techniques on social media data under specific assumptions. Recent manual analysis using data scraped from Twitter during this time by Clemson University suggests that Russian twitter agents were pro-NRA \cite{M18}. We show related and extended results in this paper.

To simplify our presentation, we take as axiomatic that there \textit{was} an operation to influence the 2016 Presidential election and construct a set of applied statistical techniques that elucidate hidden features in the data set that we assert characterizes this operation. In particular, we use natural language processing, manifold learning and Fourier analysis to show (i) emergent language communities within this sample of a troll population, (ii) the transnational scope of the operation and (iii) operational characteristics of trolls that can be used for future identification.\section*{Results}
\begin{figure}
  \centering
  \includegraphics[scale=0.4]{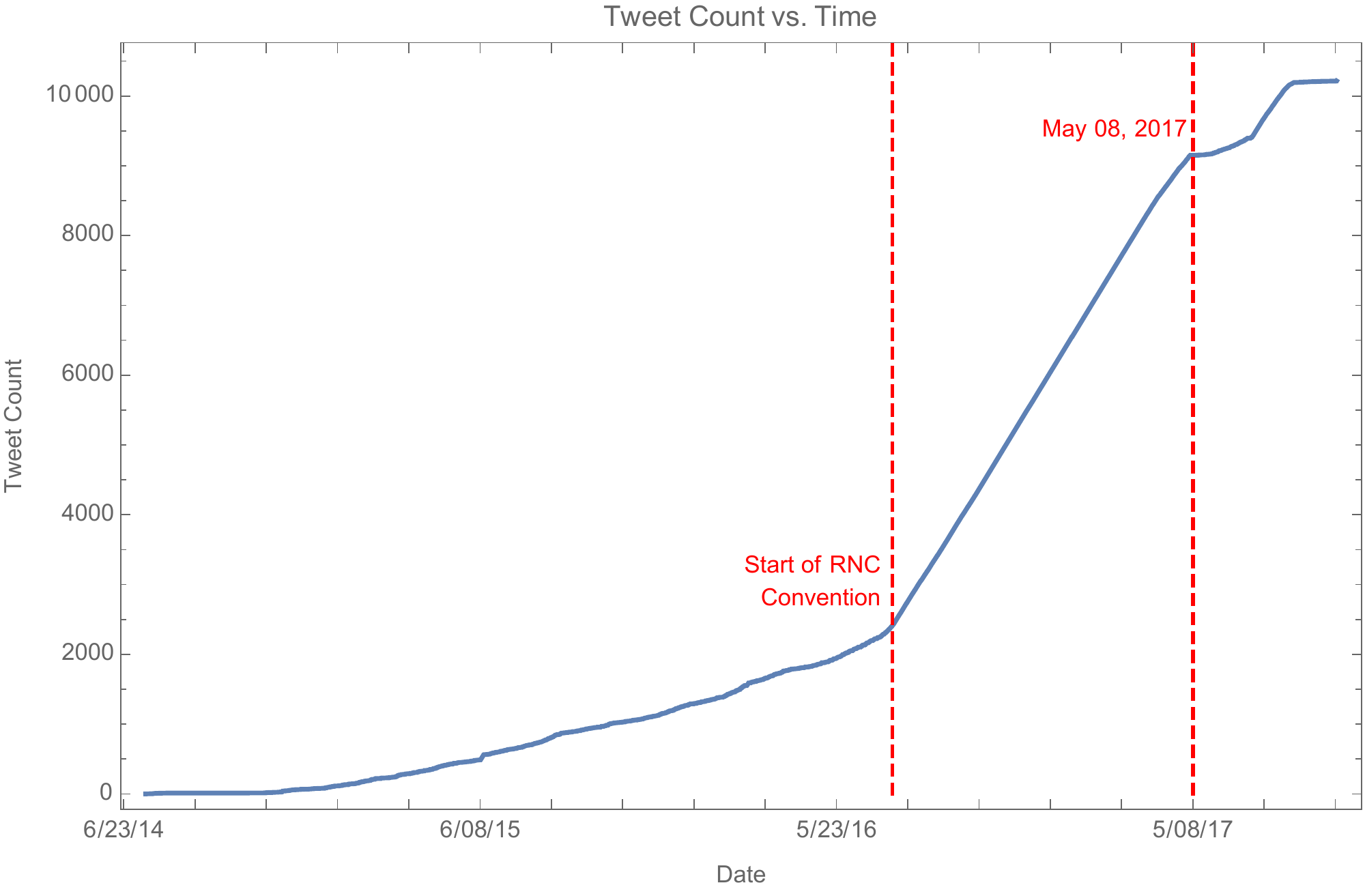}
  \caption{An annotated activity plot illustrates a change point in Russian twitter behavior corresponding to the start of the RNC convention in 2016 and immediately prior to the firing of James Comey on May 9, 2017.}
  \label{fig:Time}
\end{figure}

\begin{figure}
  \centering
  \includegraphics[scale=0.4]{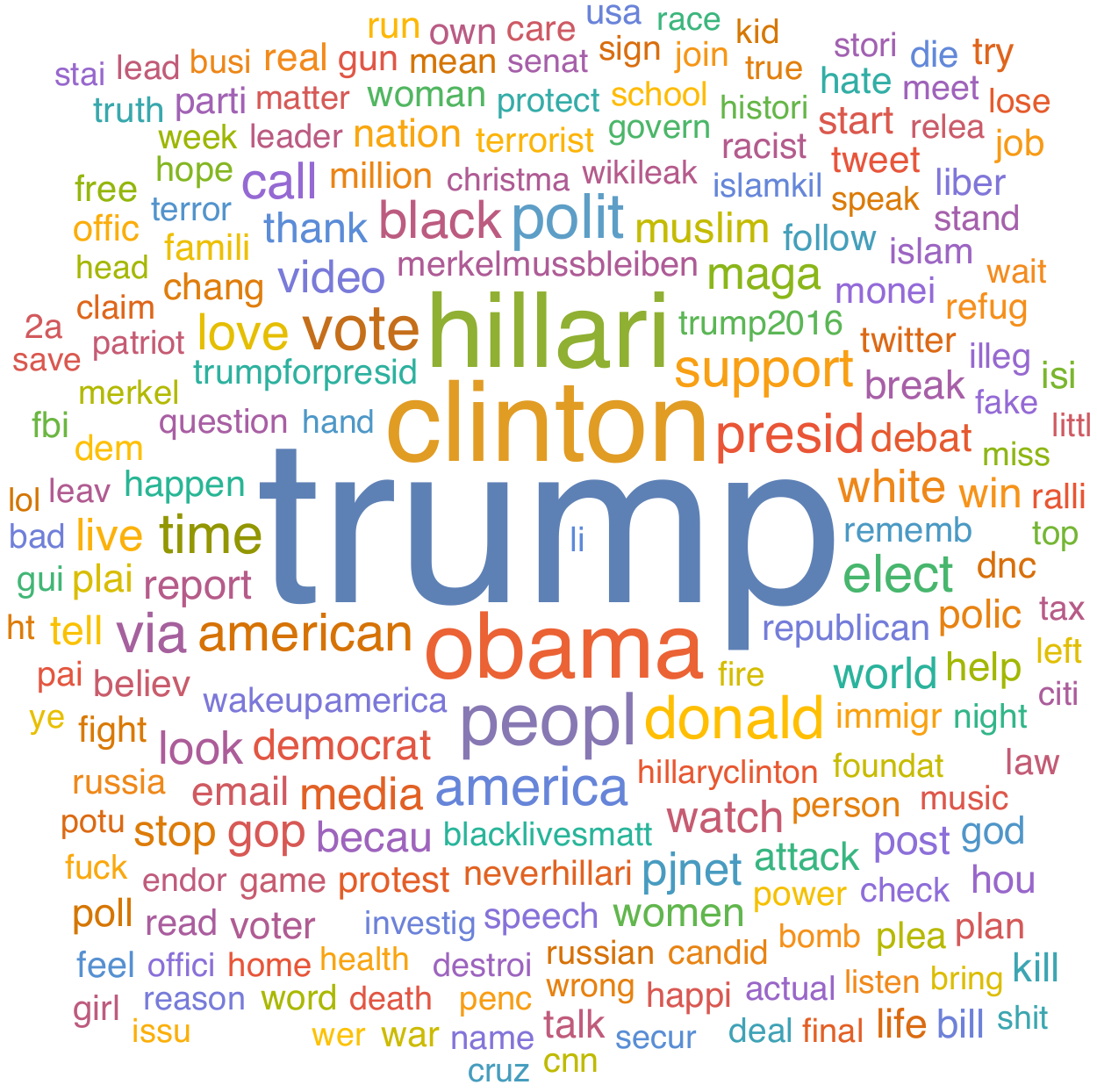}
  \caption{The most common words in the Twitter data set are shown in a word cloud. The major theme of this data set is the 2016 US Presidential election. Note, words are stemmed and normalized.}
  \label{fig:WordCloud}
\end{figure}

As noted by NBC News \cite{P18}, the Russian twitter trolls emerge as early as 2014. However, analysis of the tweet pattern using temporal meta-data reveals a structured, methodical operation that represents a type of forensic fingerprint. This operation is both reactive to \textit{ad hoc} contemporary events and targeted toward specific audiences or known scheduled events. Figure \ref{fig:Time} shows total tweet count on the ordinate and date on the abscissa illustrating a change point in operational tempo corresponding to the start of the RNC convention in 2016 and immediately prior to (or possibly concurrent with) the firing of James Comey on May 9, 2017.
Further analysis of the most common words suggests an obvious focus on the 2016 US Presidential election as illustrated in Figure \ref{fig:WordCloud}. (Note, words are stemmed and normalized.)
The fact that (then) candidate Trump is central in the world cloud is not suggestive of specific intent, but rather frequency of that particular term. As a polarizing candidate, Mr. Trump was a major news topic. However, further analysis of the data set shows specialized language features and language communities of users.

\subsection*{Language Community Analysis}
\begin{figure}
  \centering
  \includegraphics[scale=0.4]{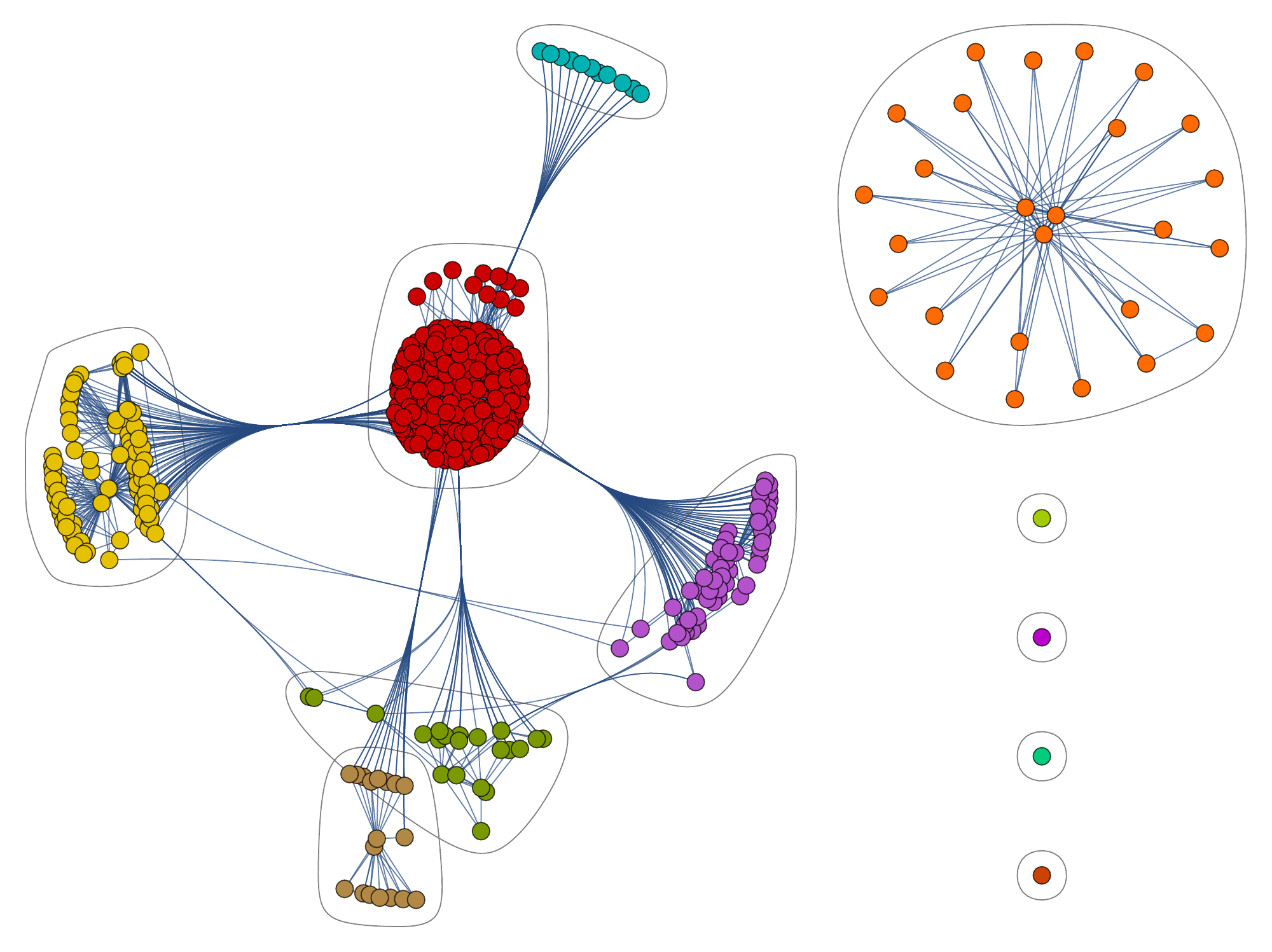}
  \caption{Eleven user communities are shown consisting of users who are tweeting about similar topics.}
  \label{fig:CommunityGraph}
\end{figure}

\begin{figure*}
  \centering
  \includegraphics[scale=0.7]{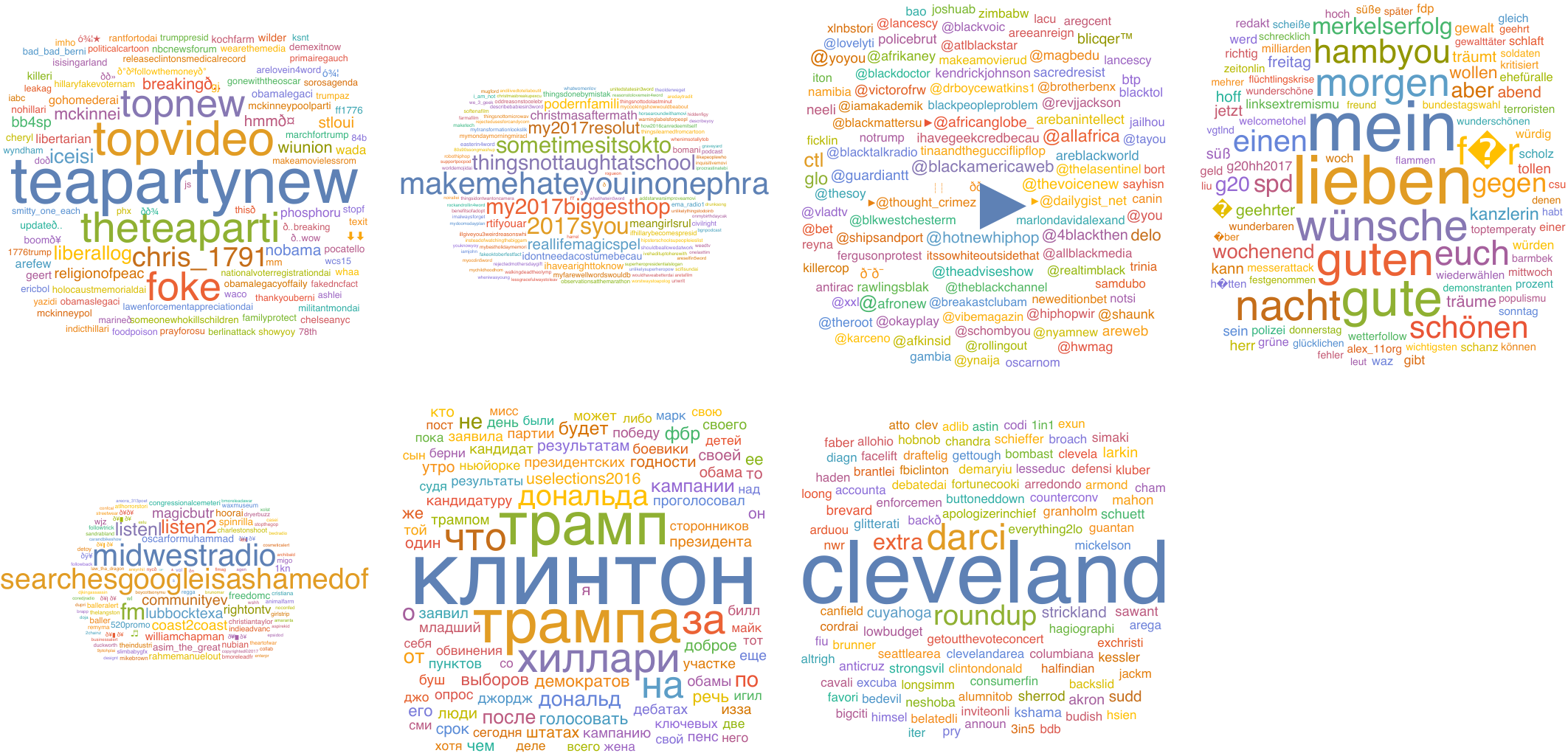}
  \caption{There are 7 major topical areas whose contents vary from specific information focused on US Tea Party Movement, general Twitter memes, topics relating to African Americans, a focus on Cleveland Ohio, one topic entirely in German and a second entirely in Russian. }
  \label{fig:WordClouds}
\end{figure*}

Treating each user as producing a \textit{document} of tweets we  used unsupervised natural language processing techniques to cluster users into specific topical communities. Common words (dynamic stopwords) were removed to allow this analysis. (Our approach is documented in both the materials and methods section and the supplemental information.) We identified 11 user communities made of the 453 users that tweeted enough to allow for inclusion in the  analysis. This is shown in Figure \ref{fig:CommunityGraph}.
Ignoring the four singleton communities, the topics discussed within the seven remaining user communities is shown in Figure \ref{fig:WordClouds}.
Read from left-to-right and top-to-bottom, the first word cloud (top left) is focused entirely on the United States Tea Party Movement. The second word cloud seems to be a junk word cloud (common twitter memes). Word cloud 3 is contains topics related to African Americans. Word clouds 5 and 7 specifically focus on the American midwest with a specific focus on Cleveland and the swing state Ohio (Word Cloud 7). Word clouds 4 and 6 are non-English (German and Russian, respectively). The inclusion of non-English topics was surprising and the unsupervised clustering was isolated these topics with no training.

Since our natural language processing engine, was trained on English, we did not remove German or Russian stopwords (or stem). Translation (using Google Translate) of the top 50 German words provided little information, while the top Russian word were more intriguing; these are shown below:
\begin{description}
    \item[German:]\hspace*{2em} my, love, good, good, wishes, night, tomorrow, for, you, hambyou, beautiful, one, against, spd, merkelserfolg, but, g20, weekend, chancellor, want, dreams, dear, evening, dreams, friday, now, hope, may, sweet, gentleman, great, being, violent, g20hh2017, leftsextremismu, gives, werd, would, right, scholz, fdp, redakt, schlaft, ehefüralle, waz, dignified, money, wow, police

    \item[Russian:]\hspace*{2em} Clinton, trump, trump, for, Hillary, what, on, Donald, o, not, on, donald, from, after, fbi, will, campaign, speech, democrats, vote, election, I, the same, her, him, uselections2016, states, than, then, morning, time, his, results, he, one, people, good, good, obama, who, president, declared, trump, obama, section, his, points, voted, presidential, victory
\end{description}
We note there were some character transliteration issue encountered, occasionally ``trump'' to be rendered ``tramp'' in translation and ``fbi'' to be rendered ``fbr.'' This may be a function of UTF-8 parsing. (See the Materials and Methods section.) We provide a further breakdown of German language tweets in the supplemental information that shows its characteristics more completely.

We chose to isolate the two communities that generated Tea Party specific materials and German Language materials to determine their language leaders. The ten most degree central vertices for the Tea Party topic are: \texttt{ameliebaldwin}, \texttt{hyddrox}, \texttt{garrettsimpson}, \texttt{patriotblake}, \texttt{jeffreykahunas}, \texttt{ten\_gop}, \texttt{thefoundingson}, \texttt{lazykstafford}, \texttt{michellearry} and \texttt{tpartynews}. We note that \texttt{ten\_gop} is currently featured in the Mueller investigation \cite{K18}. Within the Tea Party topic, the most central users were \texttt{ameliebaldwin} and \texttt{hyddrox} who posted more than any other users within this data set (see Figure \ref{fig:Behavior} in the sequel).

In German, there are three highly degree central vertices: \texttt{margarethkurz}, \texttt{klara\_sauber} and \texttt{sternandreas404}. (Illustrations of these subgraphs and the central vertices are provided in the supplemental information.) These users will re-emerge in further election analysis.

\subsection*{Transnational Election Influence}
\begin{figure}[htbp]
  \centering
  \subfloat[French Language Date Histogram]{\includegraphics[scale=0.5]{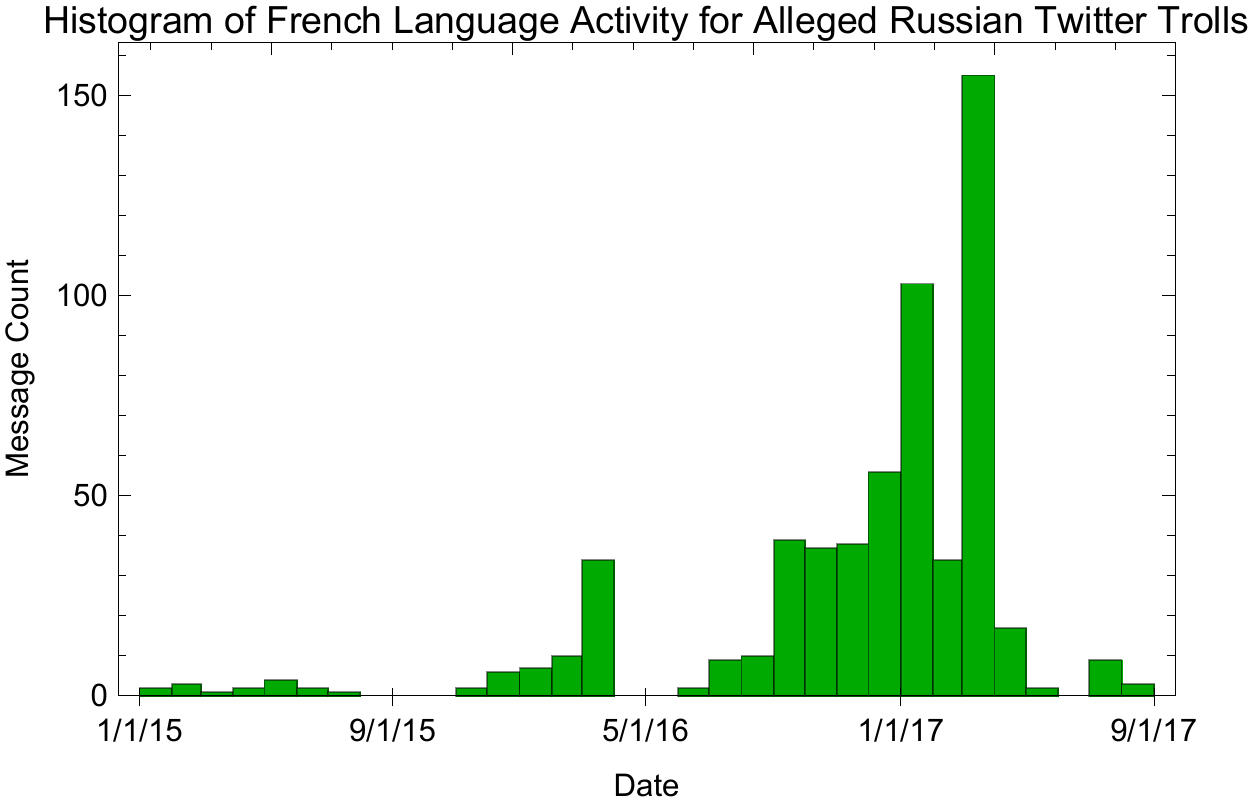}}\\
  \subfloat[German Language Date Histogram]{\includegraphics[scale=0.5]{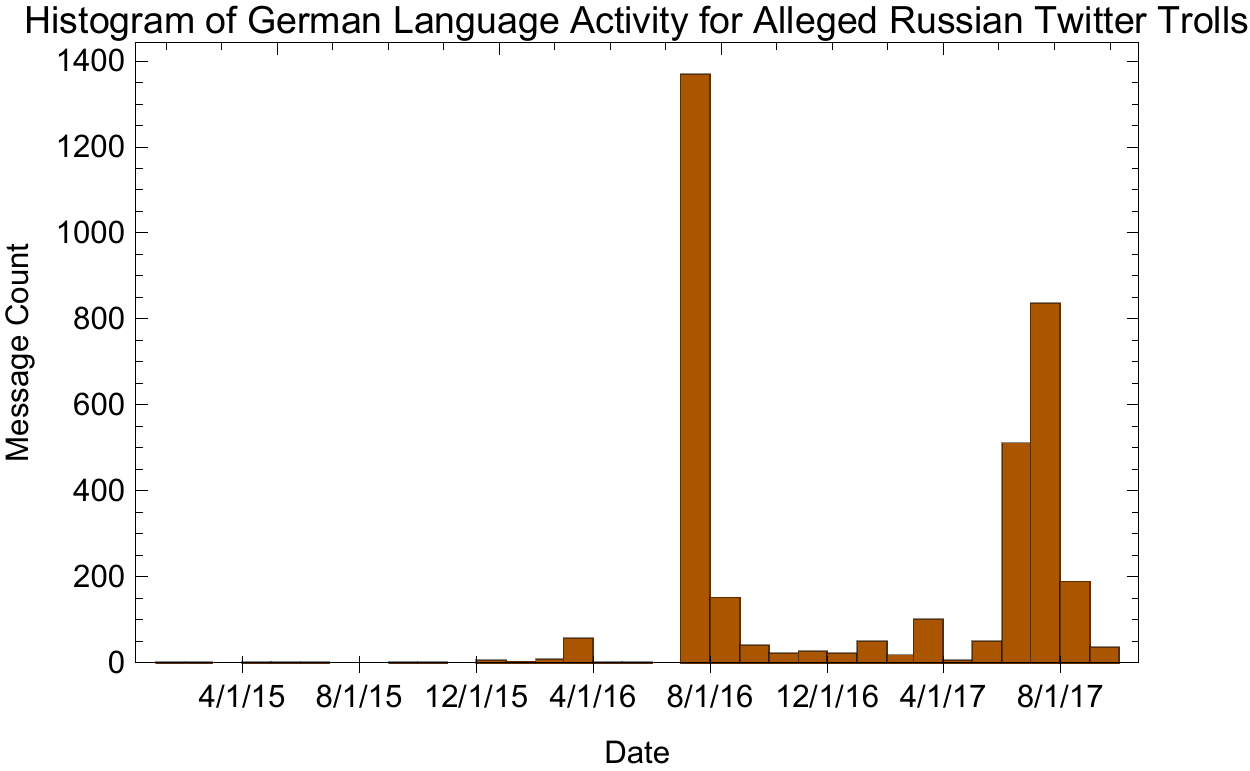}}
  \caption{French and German Language usage increase substantially around both national and local elections of importance, suggesting an influence operation that spanned multiple election cycles and nations.}
  \label{fig:LanguageHistogram}
\end{figure}

The presence of substantial non-English language fragments in the tweets suggested a broader agenda. We used an automatic language detector to identify the most likely language for each tweet and then isolated the languages used by each user. French and German tweets spiked around (i) the French National Election in 2017 and both the German Federal Elections in 2017 \textit{and} the Berlin Elections in 2016. This is shown in Figure \ref{fig:LanguageHistogram}.
In all, there were only 588 messages identified as French (these were hand verified), while there were 3522 messages in German. This explains the lack of a French word cloud and user set in the previous section. It also suggests the possibility that additional Twitter users exist separate from this data set who were responsible for influencing the French populace, though this is speculative. The relevant French election dates were April 23 and May 7, 2017. We note that this may also explain the behavioral change point in Figure \ref{fig:Time} near May 7, 2017. The Berlin State Election was held on September 18, 2016 and the Federal Elections were held on September 24, 2017. The Berlin State Election was significant because Chancellor Merkel's party was beaten by right-wing populists \cite{O16}. In the supplemental information, we analyze the content of tweets surrounding the two elections separately and show distinct differences in their content.


\subsection{Key Election Users}
We used a simple frequency count to extract the most vocal Twitter users during both the 2016 US Presidential Election and the German Federal Election. The relevant time frame of the US Presidential Election was defined to be July 18, 2016 to November 10, 2016. The relevant time frame of the German Federal Election was defined to be June 21, 2017 - September 2, 2017; the latter is based on empirical observation while the former is based on the time from the start of the Republican National Convention to two days after the election.

We defined a Twitter user to be heavy during the US Election period if (s)he posted more than 500 times. A Twitter user was defined to be heavy during the German Election period if (s)he posted more than 100 times (reflecting the smaller quantity of German message traffic). We were not able to analyze the French elections in any detail due to a lack of messages. Seven users were identified as heavy participants during the run-up to the 2016 US Presidential election: \texttt{hyddrox}, \texttt{ameliebaldwin}, \texttt{kansasdailynews}, \texttt{newspeakdaily}, \texttt{thefoundingson}, \texttt{ten\_gop}, and \texttt{dailysanfran} (read top-to-bottom and left-to-right). Notice a subset of these users were identified as linguistically central in the Tea Party topic. Others, are more geographically focused.

\begin{figure}
  \centering
  \includegraphics[scale=0.5]{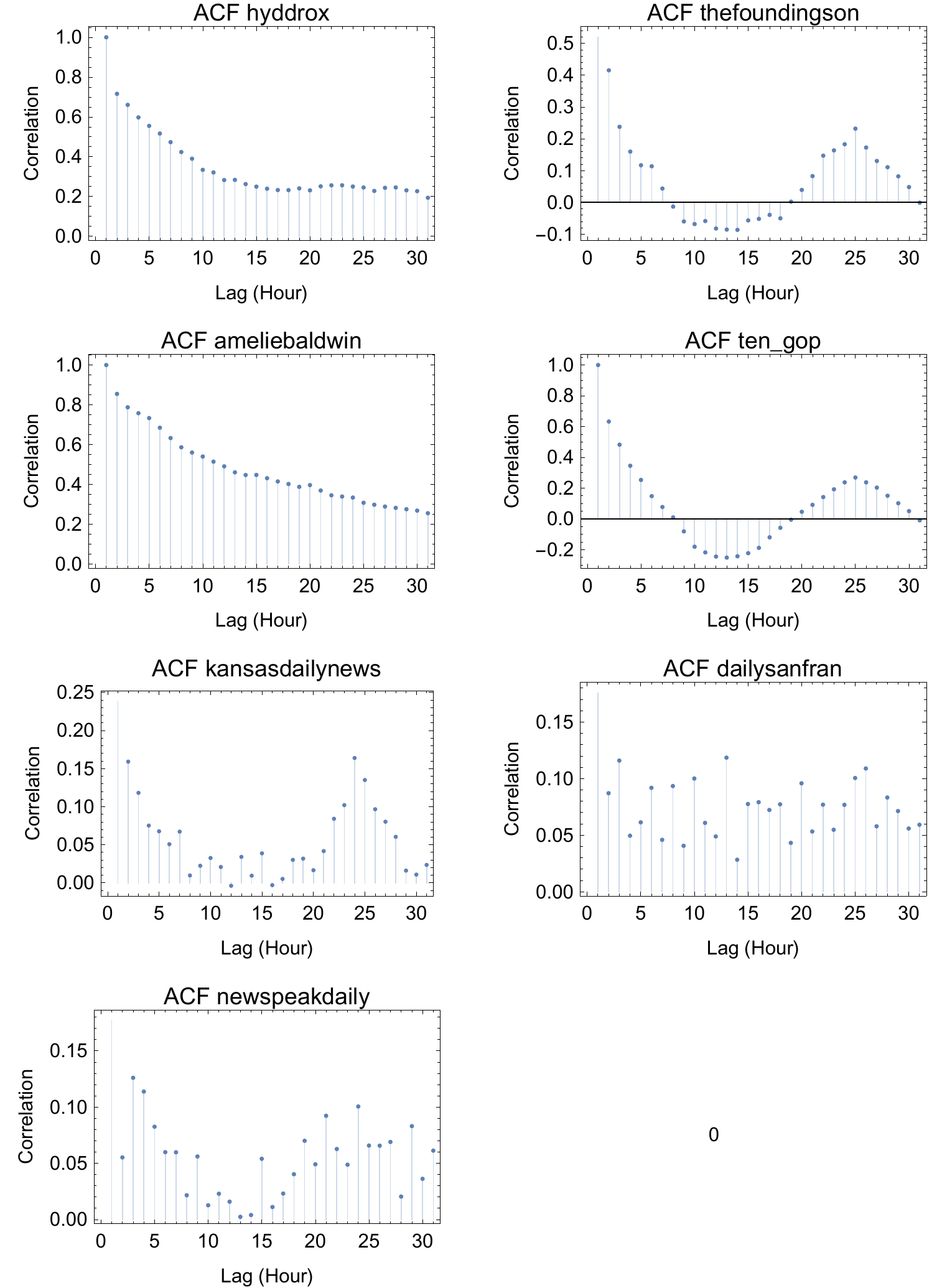}
  \caption{Autocorrelation functions of Twitter users who posted heavily during the run-up to the US election. Some users exhibit persistence or seasonal (periodic) behaviors.}
  \label{fig:USACF}
\end{figure}

To analyze the posting behavior of heavy users during the US election, we aggregated their posts into hours and computed the time series autocorrelation function (ACF). For users who where highly active during the run-up to the US election, this is shown in Figure \ref{fig:USACF}.
We note that both \texttt{hyddrox} and \texttt{ameliebaldwin} exhibit high levels of persistence in their ACF, corresponding to high tweet volumes and a high-degree of hour-on-hour volume correlation. In contrast, \texttt{ten\_gop}, \texttt{thefoundingson} and \texttt{kansasdailynews} have a quasi-periodic ACF with period between 24 and 25 hours, suggesting an operational characteristic for the users. The remaining two users have similar (though somewhat chaotic) ACF's that also exhibit some seasonality (periodicity) at 24-25 hours. We will provide a deeper time series analysis for the US election in the next section when we apply a Fourier analysis.

\begin{figure}
  \centering
  \includegraphics[scale=0.5]{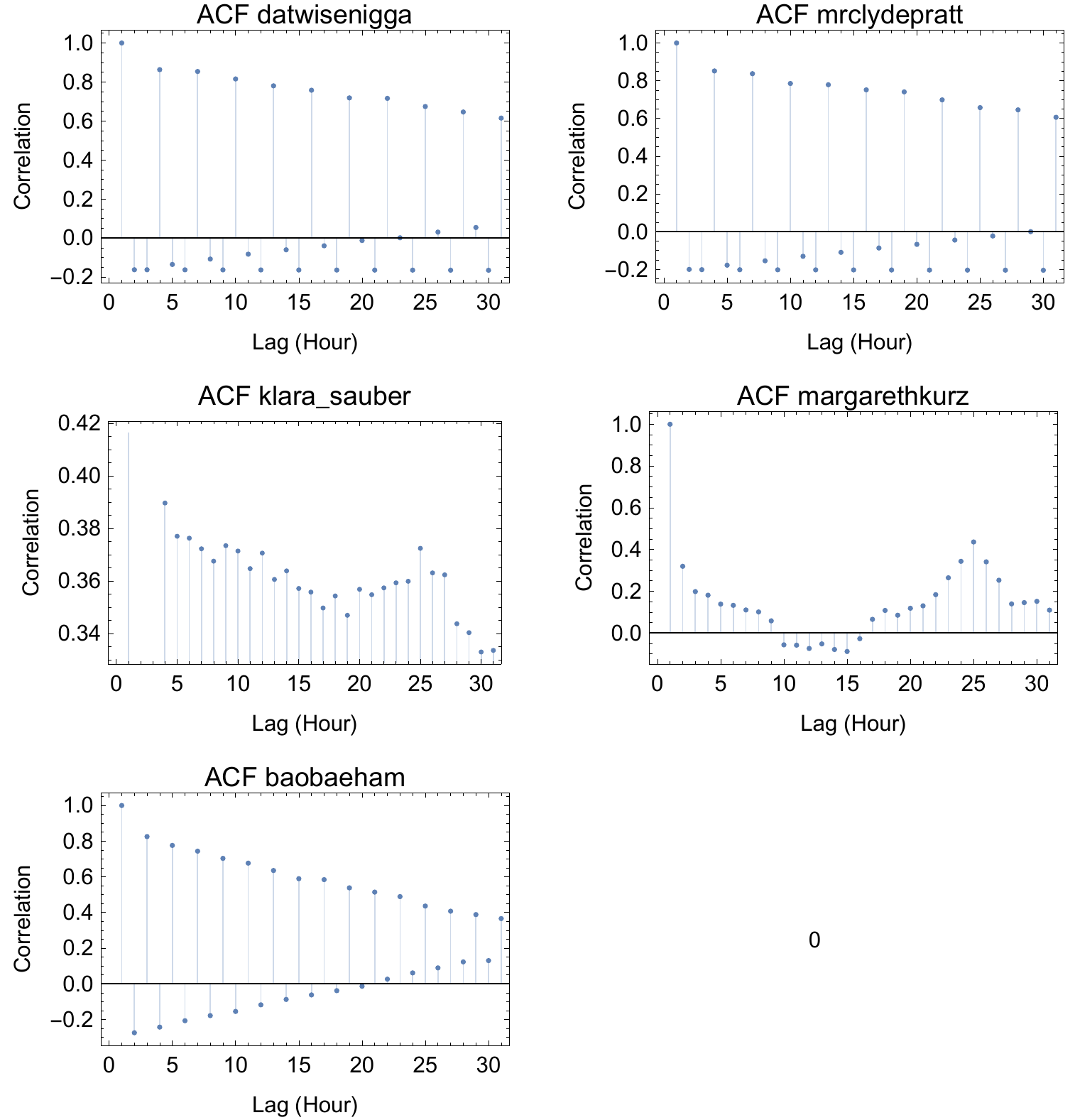}
  \caption{Autocorrelation functions of Twitter users who posted heavily during the run-up to the German election. The behavior of some users is similar to users identified in the US presidential election.}
  \label{fig:GermanACF}
\end{figure}

\begin{figure}[htbp]
  \centering
  \includegraphics[scale=0.5]{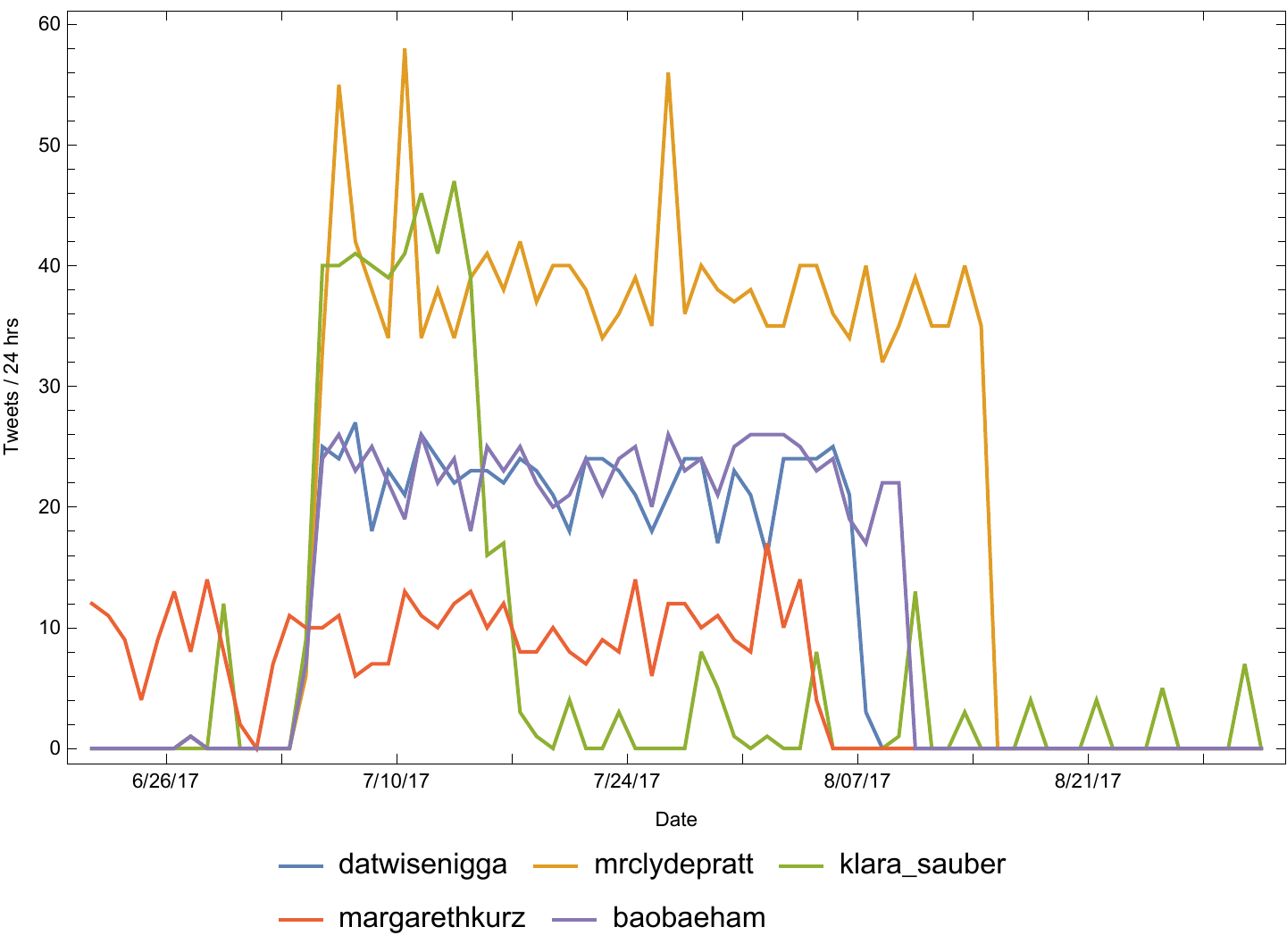}
  \caption{Time series plots of Twitter activity by frequent Tweeters in the run up to the German Federal election. Notice the entry and exit times of several users.}
  \label{fig:GermanTimeSeries}
\end{figure}

By way of comparison, Figure \ref{fig:GermanACF} shows the computed autocorrelation functions for those users identified as heavy during the German elections.
User \texttt{margarethkurz} shows the same periodic ACF at 24-25 hours as (e.g.) \texttt{ten\_gop}, which suggests a similar operational mechanism. While somewhat distinct, User \texttt{klara\_suber} shows a highly persistent ACF like \texttt{ameliebaldwin}. The remaining users who posted heavily in the run-up to the German election exhibit unusual ACF's that have a periodicity. The time series plots for the German election are shown in Figure \ref{fig:GermanTimeSeries}.
Notice the German users (mainly) begin tweeting on or about July 1, 2017 and abruptly stop on or about August 15. (See Figure \ref{fig:LanguageHistogram} for comparison.) The results during the German Federal Election are somewhat in contrast to the Berlin State Election. In 2017, Chancellor Merkel did \textit{not} lose to the far right-wing candidate, though we note the German right-wing party \textit{Alternative for Germany} became the third most populous party in the Bundestag.

Unfortunately, the signal from the local German elections (held September, 2016) was largely overwhelmed by the US election signal. Consequently, we do not include an analysis of the local German election (see Figure \ref{fig:LanguageHistogram}(b)). Most German language tweets during this time were ``one-offs'' and no user who  participated in the German Federal Elections in 2017 was present in the local elections in 2016. Interestingly, \texttt{ameliebaldwin} did tweet in German 13 times during the German local election, despite her primary language being English; we have no explanation for this.

\subsection{Fourier Analysis and Behavioral Clusters}
\begin{figure}[htbp]
  \centering
  \includegraphics[scale=0.38]{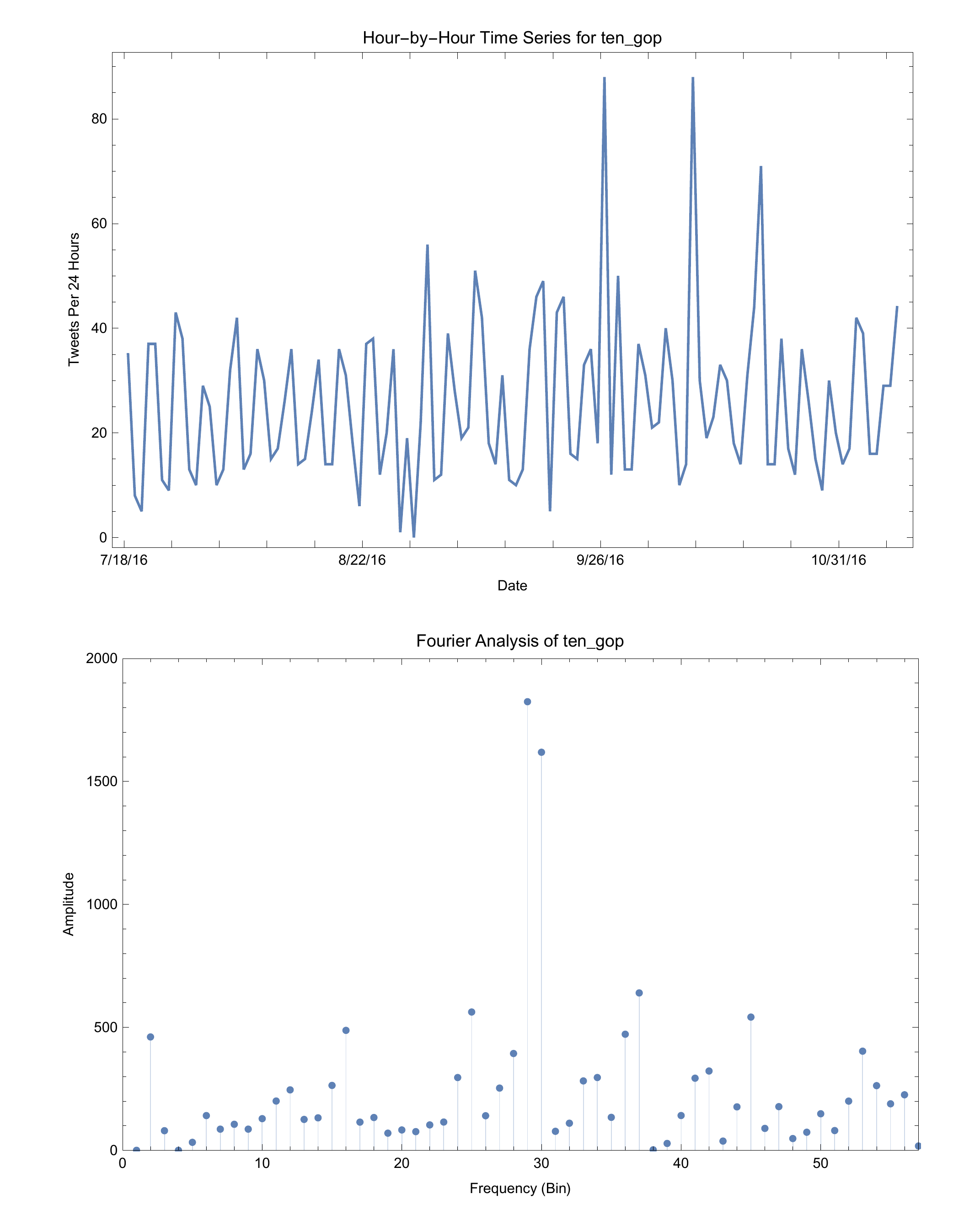}
  \caption{Time series and Fourier transform for user \texttt{ten\_gop}. Visual inspection suggests that \texttt{ten\_gop} is a strongly periodic user. This is confirmed in the spectrum. }
  \label{fig:tgop}
\end{figure}

\begin{figure}[htbp]
  \centering
  \includegraphics[scale=0.6]{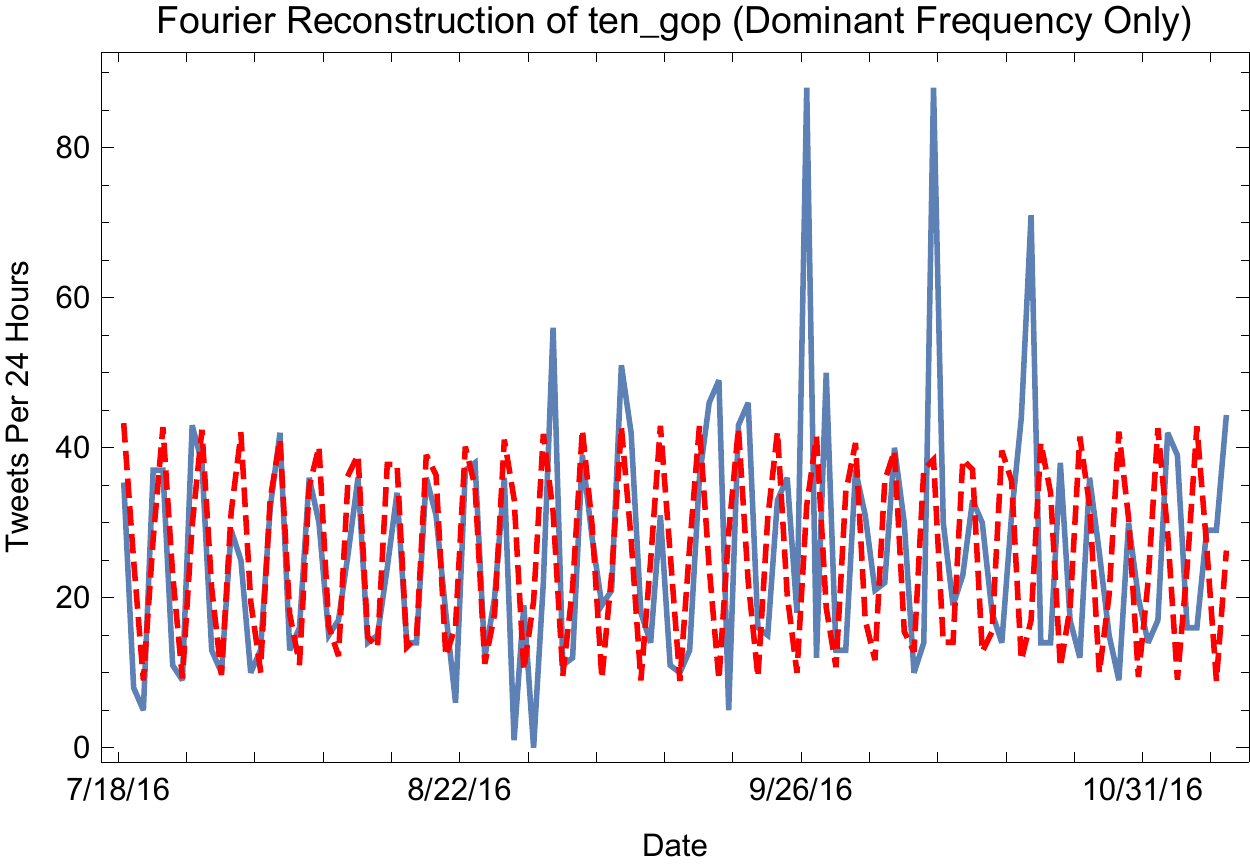}
  \caption{Reconstruction of the time series for user \texttt{ten\_gop} using the dominant frequency in his/her spectrum.}
  \label{fig:Recon}
\end{figure}

The similarity of the ACF functions identified in Figure \ref{fig:USACF} suggests common behavioral features that may be extracted and used as a profile to identify other related Twitter users. To reduce noise we converted each hourly tweet count into a 24 hour tweet count using a sliding window \textit{without overlap}. We then performed a Fourier analysis on the mean corrected signal\footnote{Mean correction removes the signal DC component allowing for enhanced analysis.} to determine whether any dominant frequencies existed within the data sample. By way of example,  consider the user \texttt{ten\_gop} whose day-by-day tweet activity and corresponding Fourier spectrum are shown in Figure \ref{fig:tgop}.
Visual inspection of the time series plot for user \texttt{ten\_gop} suggests a highly periodic signal. The corresponding Fourier spectrum shows a high amplitude component at Bin 29, corresponding to a signal with period (approximately) 3.9 days with both upper and lower harmonics. These harmonics contribute to the more irregular nature of the signal. We can confirm the validity of the spectral analysis by reconstructing the signal from its dominant frequency. This is shown in Figure \ref{fig:Recon}.
Because there are also major harmonics at bins 37 and 45, we can see quasi-periodic behavior that emerges (roughly) every 3 - 4 days in this user over the course of the run-up to the US election. Like the prior ACF analysis on hour-by-hour tweet volumes, this suggests an operational tempo that is influenced by the 24 hour news cycle, but also some exogenous influence native to the user's (i.e., \texttt{ten\_gop}'s) environment.

\begin{figure*}
   \centering
   \includegraphics[scale=0.28]{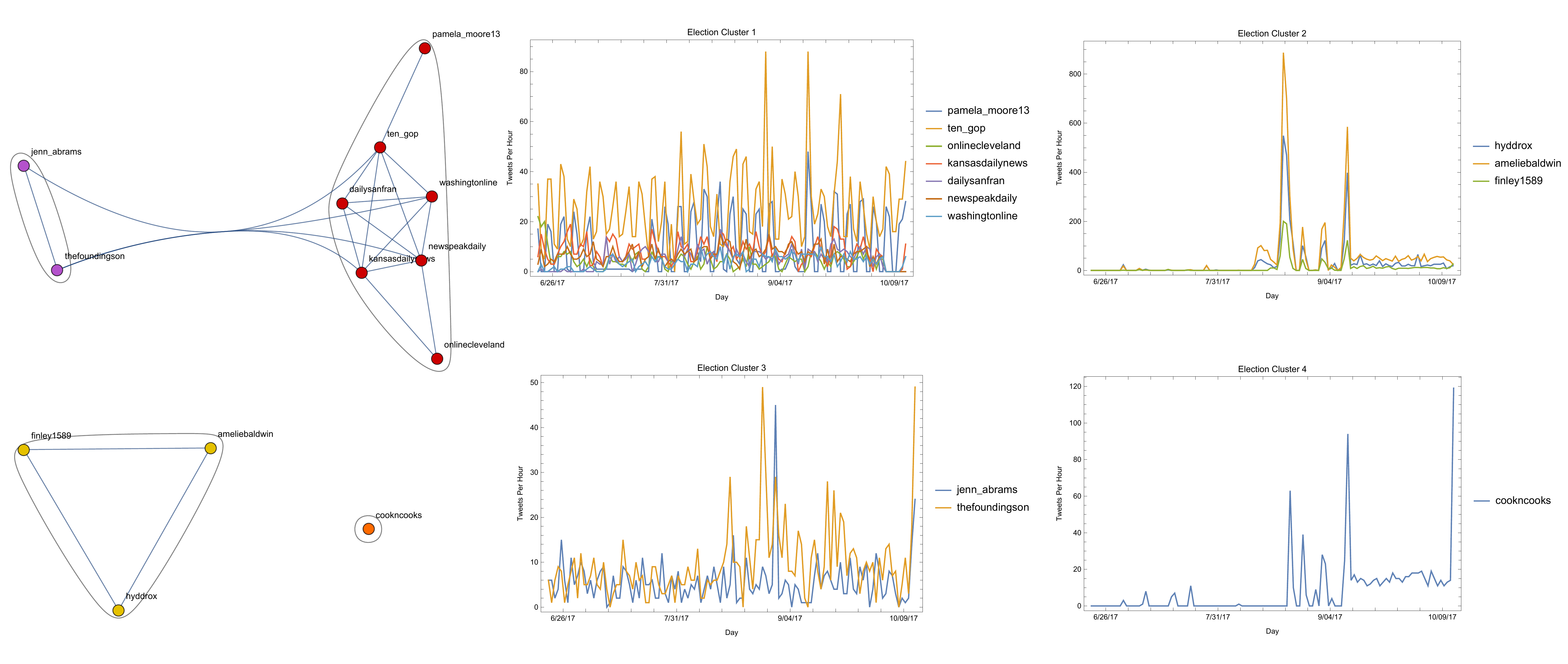}
   \caption{Using manifold learning, individual troll behaviors can be clustered together suggesting certain common operating characteristics and also suggesting a method for further detection of troll variants.}
   \label{fig:Behavior}
 \end{figure*}

 The forgoing research suggested the possibility of clustering trolls based on their behavior. To test this, we created a larger user sample set for the run-up to the US presidential election by considering all users who posted more than 400 times. We then clustered the individuals using their spectra using a modification of the Isomap approach \cite{TSL00} and Multidimensional Scaling \cite{BG97} to manifold learning \footnote{Manifold learning is (essentially) non-linear data embedding and clustering on a high-dimensional manifold, rather than in Euclidean space.}\cite{BN02}. Clustering on spectra is considered in \cite{FGNP+03} where the authors use principal components analysis to project and cluster acoustic spectra for use in vehicle classification by acoustic sensor networks; this technique is inspired from that work. The resulting user clusters and comparative plots of their tweet time series are shown in Figure \ref{fig:Behavior}.
 The clustering shows common behaviors among the thirteen higher user posters during the run-up to the election.

\section*{Discussion}
Twitter and other social media analysis is an extensively studied area due to the relative simplicity of extracting large volumes of data. By way of example, \cite{BWJT17,LSD12,CHBG10,WHMW11,ZJHS11,KJKP13,SSG13} provide details on alternate twitter topical analysis, data extraction methods and influence measures. To our knowledge, this is the first account of the Russian troll data using advanced methods from applied statistics.

There are several relevant features that have been identified in the results that warrant discussion. In Figure \ref{fig:Time}, the change in operational tempo coinciding with the start of the 2016 Republican convention suggests a highly targeted operational goal, intended to exploit the heightened level of  post-convention buzz on Twitter. The second change demonstrates a responsive operational tempo and may be news-cycle driven\footnote{James Comey was terminated on May 9, 2018.}. It is clear the tempo of operations picks up again a few days later, at which point the data set terminates.

The topic of the data set is clearly the US Presidential Election as seen in Figure \ref{fig:WordCloud}, however using techniques from document clustering and automated language detection, we can support the hypothesis that this data set contains more than just troll activity surrounding the US presidential election of 2016 (see Figures \ref{fig:WordClouds} and \ref{fig:LanguageHistogram}). This leads to three new questions: (i) Why re-use Twitter handles across multiple nations? (ii) Is this truly the entire set of Russian twitter trolls? and (iii) If not, who were the others? It is also worth noting that 11.9\% of all users sent messages in English, French and German, while 15.6\% of all users sent messages in both English and German. More tellingly, 75\% of \textit{all messages} that were sent were sent by these polyglots. However, the highly active users during (e.g.) the German elections spoke only German. There is no convenient way to resolve these characteristics, though it does suggest a contagion approach in which support for far-right candidates may be being urged across multiple nations simultaneously in order to create a global sense of change. Unfortunately, this is speculative and requires further data to confirm or refute this hypothesis. We also note, this may be a function of the data sample itself. Since this data was provided by Twitter to NBC News for a specific purpose, it's clear it is not a randomly sampled set.

The ability to answer Question (ii) and (iii)  above (Is this all the trolls, are there others?) we considered the possibility that there might be a set of common operating characteristics in the tweet tempo of the trolls. This hypothesis was based on evidence from Figures \ref{fig:Time} and \ref{fig:GermanTimeSeries} that there was a a structured, methodical operation in place to activate the trolls and the suggestive autocorrelation functions (see Figures \ref{fig:USACF} and \ref{fig:GermanACF}), which showed high levels of periodicity in the time series. Our Fourier methods, inspired by \cite{FGNP+03}, show that there are (at least) four operational tempos leading to distinct, but classifiable troll behaviors. With user \texttt{cookncooks}, we note that (s)he became more active toward the end of the 2016 US Presidential election cycle and was highly active during the French election cycle (not shown), though tweeted \textit{mainly} in English. This may explain his isolation to a singleton cluster. The remaining clusters do exhibit similar behaviors in their time series. (Note, only the spectra were used for clustering). The larger group containing \texttt{ten\_gop} could be further sub-clustered, though visual inspection suggests that group has a similar set of operating characteristics with varying amplitudes.

To summarize: in this paper we found evidence to support the hypothesis that there was a structured, methodical operation that represents a type of forensic fingerprint in operation around the Twitter provided Russian twitter trolls. Using natural language processing and advanced statistical methods we showed (i) emergent language communities within the troll population This led us to identify (ii) the transnational scope of the operation, which led to new questions on the existence of other trolls not identified. To answer this question, we developed techniques to classify (iii) operational characteristics of trolls that can be used for future identification. This supports the hypothesis that these trolls can be detected (at least partially) automatically, especially if a seed troll (or trolls) are identified. The identification process \textit{does not} need to rely solely on text content, which can be difficult to separate. The proposed methods should generalize and do not rely on content specific information, unless the specific language stopwords or stemmers (where appropriate) are used.

\subsection*{Conclusions and Future Directions}
In this paper, we analyzed a data set purported to represent Russian troll twitter messages and showed that an operational fingerprint could be identified. We also showed that the twitter users separated into 11 operational clusters based on language and illustrated methods to detect new Twitter trolls based on their operating characteristics using Fourier analysis.

Unfortunately, as attempts at using social media continue, this work is only the beginning. There is simply too much data volume for this to be done by hand; it is impossible to rely on any team of investigators to sift through millions of tweets by hand in order to stem future influence operations. For future work, one could evaluate the use of these algorithms in an online setting. Ideally a data set of both influencer troll and non-troll posts could be collected to determine the effectiveness of separating out these two classes of users. It is important to note that the operational characteristic hypothesis relies on the assumption that there is a operational management structure on the troll backend. These techniques may not work with ordinary Internet trolls (in the classic sense), since they may not necessarily present a methodical operational fingerprint. Additional tests to confirm these assertions should also be considered.

Most importantly, in order to effectively apply the techniques outlined in this paper, it is important to be able to apply these techniques with some level of certitude that the ``foreign troll activity'' being detected, is, in fact just that. There is a high level of confidence that these techniques can be developed into operational capabilities and used to to associate transmission profiles to future Russian Twitter trolls. There is also confidence that these profiles could be used to monitor for the appearance of higher and higher levels of transmission intensity that are suggestive of the tactics used by the foreign trolls. This would enable legitimate government authorities to identify foreign influence and meddling. However, perhaps of higher criticality would be the ability to distinguish between inappropriate foreign meddling and legitimate national political organizations exercising their 1st Amendment rights during election season. These legitimate organizations might also employ structured, methodical procedures to effectively message their intended audiences.  Potentially, a broader holistic perspective that evaluates whether the sources of the twitter messages are demonstrating multi-lingual characteristics that do not match the anticipated or usual expected local area languages and enriched with other meta-data that could reveal the geo-location of the twitter sources. When combined with some level of multi-national cooperation among democratic nations to monitor bad actors, this might mitigate those risks of misidentification, and help ensure it is possible to distinguish friend from foe without negatively impacting freedom of expression.

\section*{Materials and Methods}
Data were obtained from NBC News \cite{P18} and consisted of two comma separated value (CSV) files of tweets and users. The files were both UTF-8 encoded. We parsed 203,451 tweets and identified 453 users with useable tweets. There were 514 users noted in the data set. We observed minimal parsing problems related to the encoding, though this may have been a function of the software system. All statistical analysis and image generation was done using Mathematica (version 11.3). Some word counting and text comparison was done in customized C++ code. English words were stemmed and stopwords were removed using a custom stopword list. For troll community detection (see Figures \ref{fig:CommunityGraph} and \ref{fig:WordClouds}), we removed \textit{data set stopwords} defined as any word that is used by more than $33\%$ of the users. This allowed language community structure to emerge naturally.

Clustering for both the user-topic analysis (see Figure \ref{fig:CommunityGraph}) and behavior analysis (see Figure \ref{fig:Behavior}) were done using maximum modularity clustering \cite{N06,LF09} on an inferred graph structure using internal Mathematica commands. This extends the ordinary Isomap \cite{TSL00} and Multidimensional Scaling approach \cite{BG97} to manifold learning. Edge weights were determined by appropriate similarity measures of relevant feature vectors. Specific algorithms are discussed in the supplemental information.

\appendix
In this supplemental information, we provide details on the manifold learning approach used to generate the user clusters based on both language use characteristics and tweet production characteristics.

\section*{Generating Clusters Based on Language}
Algorithm \ref{alg:LangCluster} is used to generate the user clusters based on language (see next page). It is a compilation of several pre-existing methods in natural language processing \cite{MS99} and manifold learning \cite{TSL00,BN02} with a novel variant that makes use of graph analytics developed in the physics community \cite{N06,N16,N13}.

We assume some minimal knowledge from statistical language processing. In particular, a \textit{stopword} is a word with little value to text meaning; examples in English include ``the'' and ``a.'' Word stemming removes common endings to ground the text into a more systematic vocabulary. For example, ``ending'' is replaced with ``end.''

We assume the input is a sequence of tweets and users. In Line 1, the tweets from User $i$ are concatenated to form a synthetic document. For the remainder of this algorithm if $w$ is a word, let $\#(w,i)$ be the number of times User $i$ uses word $w$. In Line 2, the words in the document are passed through a stemmer. Since we did not know that multiple languages were in use \textit{a priori}, we used the standard English (Porter) stemmer available in Mathematica 11.3. However, multiple stemmers and language detectors could be used at this step. In Line 3, we use a custom stopword list consisting of the (stems of) most common words in a language. Again, this was specialized to English, however it is possible to generalize this step. The remaining non-stopwords were used to create term count vectors for each user in Line 4. In Line 5, a binary term-user matrix is constructed. The entries are indictors determining whether a term is used by a user. A \textit{dynamic stopword set} is initialized in Step 7. In Lines 8 - 10, we iterate through the term-user matrix and determine whether a term is used by more than $p \times 100\%$ of the users, with $p \in (0,1)$. For our study we set $p = 0.3333$. The rows corresponding to the words that were used by more than $p \times 100\%$ are added to $T$.
\begin{algorithm}[htbp]
  \begin{algorithmic}[1]
    \caption{Construct User Clusters from Linguistic Features}
    \label{alg:LangCluster}
    \Statex\textbf{Input: }Raw tweets with users
    \Statex\textbf{Result: }Clusters of Users
    \State Concatenate tweets per user to make documents
    \State Stem all words in documents
    \State Remove stopwords by their stems
    \State Generate word counts per user
    \State Create binary term-user matrix:
    \begin{displaymath}
      \mathbf{X}_{ij} \leftarrow \begin{cases}
        1 & \text{If User $j$ uses term $i$} \\
        0 & \text{otherwise}
    \end{cases}
    \end{displaymath}
    \State Define $\{n_r,n_c\}\leftarrow \mathrm{Dimensions}[\mathbf{X}]$
    \State Define $T \leftarrow \emptyset$\;
    \For{$i \in \{1,\dots,n_r\}$}
     \If{$\sum_j\mathbf{X}_{ij} > p \cdot n_c$}\Comment{\texttt{We set $p=0.333$.}}
     \State $T\leftarrow T \cup \{i\}$
    \EndIf
    \EndFor
    \State Delete all terms corresponding to indexes in $T$
    \State Set $T \leftarrow \emptyset$.
    \For{$i \in \{1,\dots,n_c\}$}
    \State Compute mean word count $\bar{w}$ and word count variance $S_w^2$ for User $i$.
    \State Fit a Gamma distribution $\Gamma(k,\theta)$ for User $i$'s word use with:
    \begin{displaymath}
        k = \frac{\bar{w}^2}{S_w^2} \qquad \theta = \frac{S^2}{\bar{w}}
    \end{displaymath}
    \For{Each word used by User $i$}
      \If{$\#(w,i)$ is in the $q\times 100\%$ percentile of $\Gamma(k,\theta)$}
      \State $T \leftarrow T \cup \{w\}$.
      \Comment{\texttt{We set $q = 0.9$. These are \textit{keywords}.}}
      \EndIf
    \EndFor
    \EndFor
    \State Delete all terms not in $T$.
    \State Create new term-user matrix: $\tilde{\mathbf{X}}_{ij}$ is the number of times User $j$ uses term $i$.
    \State Compute modified adjacency matrix: $\mathbf{A} \leftarrow \tilde{\mathbf{X}}^T\cdot\tilde{\mathbf{X}}$
    \State Compute $\tilde{\mathbf{A}}$ by zeroing the diagonal of $\mathbf{A}$
    \State Set $B_i \leftarrow k^\text{th}\text{ largest value in $\mathbf{A}_{i\cdot}$}$
    \Comment{\texttt{The previous line creates a bound; in practice, we set $B_i$ to the 3rd largest weight in row $i$ denoted $\tilde{\mathbf{A}}_{i\cdot}$. The next for-loop thins the graph.}}
    \For{$i \in \{1,\dots,n_c\}$}
      \For{$j \in \{1,\dots,n_c\}$}
        \If{$\tilde{\mathbf{A}}_{ij} < \min\{B_i,B_j\}$}
         \State $\tilde{\mathbf{A}}_{ij} \leftarrow 0$
        \EndIf
      \EndFor
    \EndFor
    \State Create weighted graph $G$ with weight matrix $\tilde{\mathbf{A}}$
    \State Cluster vertices using maximum modularity clustering.
  \end{algorithmic}
\end{algorithm}

After deleting the dynamic stopwords, we identify keywords across all documents in Lines 13 - 18 using frequency. While it is the case that many documents follow the Zipf-Mandelbrot law and their word counts follow a Zipf-Mandelbrot distribution \cite{Z32,M65}, we have found it useful to smooth these distributions and use a percentile cutoff. To do this, we fit a Gamma distribution to the word counts and then remove words that do not fall in the $q \times 100\%$ percentile of the distribution. For this study, we set $q = 0.9$. We chose the Gamma distribution because it has non-negative support, and being a two parameter distribution is widely adaptable. We note, there are alternate way of identifying keywords \cite{MS99}.

After removing non-keywords, the term and user set is fixed and a new term-user matrix $\tilde{\mathbf{X}}$ can be computed where each column is the (reduced term) term counts for the corresponding user. To understand the remainder of the algorithm, we think of $\mathbf{X}$ as a sub-matrix of the weighted adjacency matrix $\mathbf{B}$ of a bipartite graph between terms and documents where:
\begin{displaymath}
    \tilde{\mathbf{B}} = \begin{bmatrix}
       \mathbf{0} & \tilde{\mathbf{X}}\\
       \tilde{\mathbf{X}}^T & \mathbf{0}
    \end{bmatrix}
\end{displaymath}
Suppose that $\mathbf{B}$ has the same structure as $\tilde{\mathbf{B}}$, but with all positive values replaced by $1$. It is a classic result in graph theory that $\mathbf{B}^2$ has in its $(i,j)$ position the number of walks of length $2$ from vertex $i$ to vertex $j$. Thus, if $\mathbf{X}$ is again the binarized form of $\tilde{\mathbf{X}}$, then:
\begin{displaymath}
  \mathbf{B}^2 = \begin{bmatrix}
  \mathbf{X} \cdot \mathbf{X}^T & \mathbf{0}\\
  \mathbf{0} & \mathbf{X}^T \cdot \mathbf{X}
\end{bmatrix}
\end{displaymath}
Here $\mathbf{X}^T\cdot\mathbf{X}$ is a square matrix whose $(i,j)$ position is the number of terms shared between User $i$ and User $j$ when $i \neq j$. Passing to the general case, the $(i,j)$ element of $\tilde{\mathbf{X}}^T\cdot \tilde{\mathbf{X}}$ is a measure of the number of terms shared in common between User $i$ and User $j$ (when $i \neq j$) weighted by occurrence counts.

In Lines 20-27, we create the adjacency matrix $\tilde{\mathbf{A}}$ by using the process described above and zero the diagonal. We then compute bounds for each User, by finding the $k^\text{th}$ largest value. (In our work we used the third largest value.) In Lines 24 - 27, we zero an edge weight (i.e., delete edges) at index $(i,j)$ if it is less than the bound computed for User $i$ or User $j$. In essence, this creates a version of a preferential attachment graph where (i) preference is based on shared language, (ii) each user  prefers to be attached to $k$ other users but (iii) User $i$ may be attached to more than $k$ users if there are more than $k$ other users who have User $i$ as one of their top $k$ connections. If modeled as a directed graph, this would be an $(n_c, k)$ directed Barabasi-Albert network \cite{AB00} with a hidden preference based on topic.

Finally, in Lines 20 - 21, we form the graph and use maximum modularity clustering defined by Newman \cite{N06} to be build user communities. This is a variation on the Isomap \cite{TSL00} and Multidimensional Scaling \cite{BG97} approaches to manifold learning.

As shown in Figure 4 of the main text, the results provide a convincing set of user communities with coherent topics. Additionally, even though the subroutines were specialized for English, this approach identifies non-English language speaking groups and topics with no additional training.

\section*{Generating Clusters Based on Behavior}
Algorithm \ref{alg:TweetCluster} is used to generate clusters of users based on tweet behavior. It is an extension of work in \cite{FGNP+03}, which used principal components analysis and $k$-means clustering to group acoustic signals into vehicle classes using their spectral properties. Since we use the same manifold learning method that was used in Algorithm \ref{alg:LangCluster}, this approach is a logical extension of the work in \cite{FGNP+03}.

\begin{algorithm}[p]
  \begin{algorithmic}[1]
  \caption{Construct User Clusters from Tweet Dynamics}
  \label{alg:TweetCluster}
  \Statex Hour-by-Hour Tweet Counts by User $\{\mathbf{x}_1,\dots,\mathbf{x}_n\}$
  \Statex Clusters of Users.
  \For{$i \in \{1,\dots,n\}$}
    \State Compute the discrete Fourier transform $\mathbf{X}_i = \mathcal{F}(\mathbf{x}_i) = \langle{a_{i_1},\dots,a_{i_m}}\rangle \in \mathcal{C}^m$
    \State Compute the power spectrum: $\mathbf{P}_i = \langle{|a_{i_1}|^2,\dots,|a_{i_m}|^2}\rangle \in \mathbb{R}^m$
  \EndFor
  \State Let $\mathbf{A} \leftarrow \mathbf{0} \in \mathbb{R}^{n \times n}$.
  \For{$i \in \{1,\dots,n\}$}
     \For{$j \in \{1,\dots,n\}$}
      \If{$i \neq j$}
      \State Define:
        \begin{displaymath}
          \delta_{ij} \leftarrow 1 - \frac{\langle{\mathbf{P}_i,\mathbf{P}_j}\rangle}{\lVert\mathbf{P}_i\rVert\lVert\mathbf{P}_j\rVert}
        \end{displaymath}
        \If{$\delta_{ij} > T$}
           \State Set: $\mathbf{A}_{ij} \leftarrow \delta_{ij}$
        \EndIf
      \EndIf
    \EndFor
  \EndFor
  \State Create weighted graph $G$ with weight matrix $\mathbf{A}$\;
  \State Cluster vertices using maximum modularity clustering.
\end{algorithmic}
\end{algorithm}

We assume that for User $i$ a time series of (hour-by-hour) tweet counts is provided in the vector $\mathbf{x}_i$. In Lines 1 - 3, we compute the discrete power spectrum of $\mathbf{x}_i$ denoted $\mathbf{P}_i$. The set $\{\mathbf{P}_i\}_{i=1}^n$ will be used to create similarity scores between the users. In Line 5 a zero matrix is created to hold these similarity scores.

In Lines 4 - 10 we construct the cosine similarity measure between User $i$ and User $j$. Since $\mathbf{P}_i \in \mathbb{R}_+^m$ (the positive orthant) for all $i \in \{1,\dots,n\}$, the computed cosine distance satisfies:
\begin{displaymath}
  0 \leq \frac{\langle{\mathbf{P}_i,\mathbf{P}_j}\rangle}{\lVert\mathbf{P}_i\rVert\lVert\mathbf{P}_j\rVert} \leq 1,
\end{displaymath}
Here, $\langle{\cdot,\cdot}\rangle$ denotes the standard inner product and $\lVert\cdot\rVert$ denotes the standard norm in Euclidean space. In Line 9, we use a threshold $T$ to make the similarity matrix $\mathbf{A}$ sparser. In our work, we use $T = 0.7$; this corresponds to requiring the angle between $\mathbf{P}_i$ and $\mathbf{P}_j$ to be less than $\approx 45^\circ$ degrees in order for the spectra to be declared similar. Having constructed $\mathbf{A}$, we build a graph structure $G$ and apply maximum modularity clustering, just as in Algorithm \ref{alg:LangCluster}.

We note that we attempted to apply standard principal components analysis to $\{\mathbf{X}_i\}_{i=1}^n$ to reduce noise in the data and improve the clustering. On this data set, we found anecdotal evidence that the clusters were less cohesive than when no noise reduction was applied. It is possible that in longer signals, noise reduction will improve clustering and we were not able to observe this because of the (relatively) short duration of the run up to the US presidential election.

\section*{German and Russian User Clusters}
\begin{figure}[p]
  \centering
  \subfloat[German Cluster]{\includegraphics[scale=0.5]{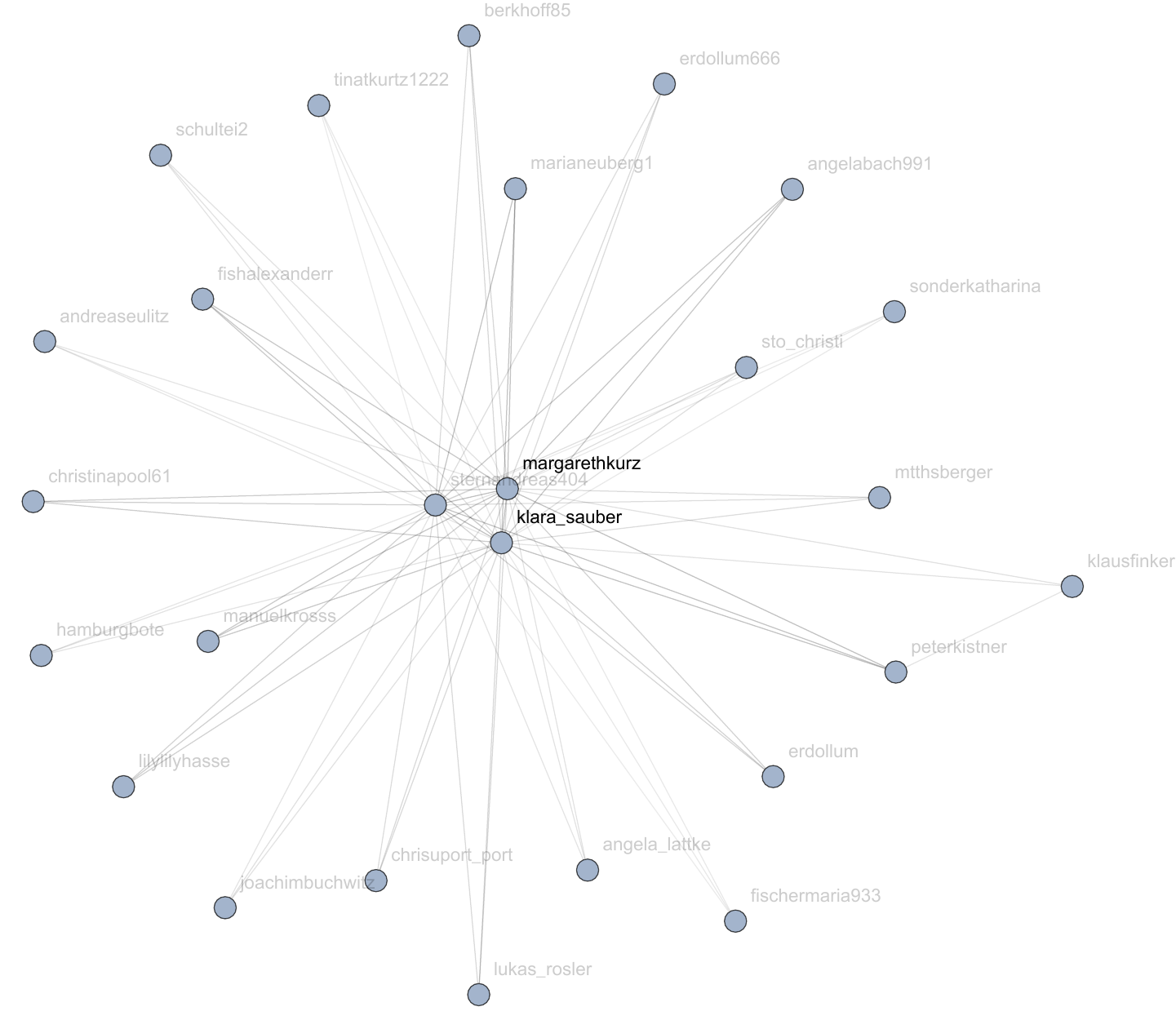}} \qquad
  \subfloat[Russian Cluster]{\includegraphics[scale=0.5]{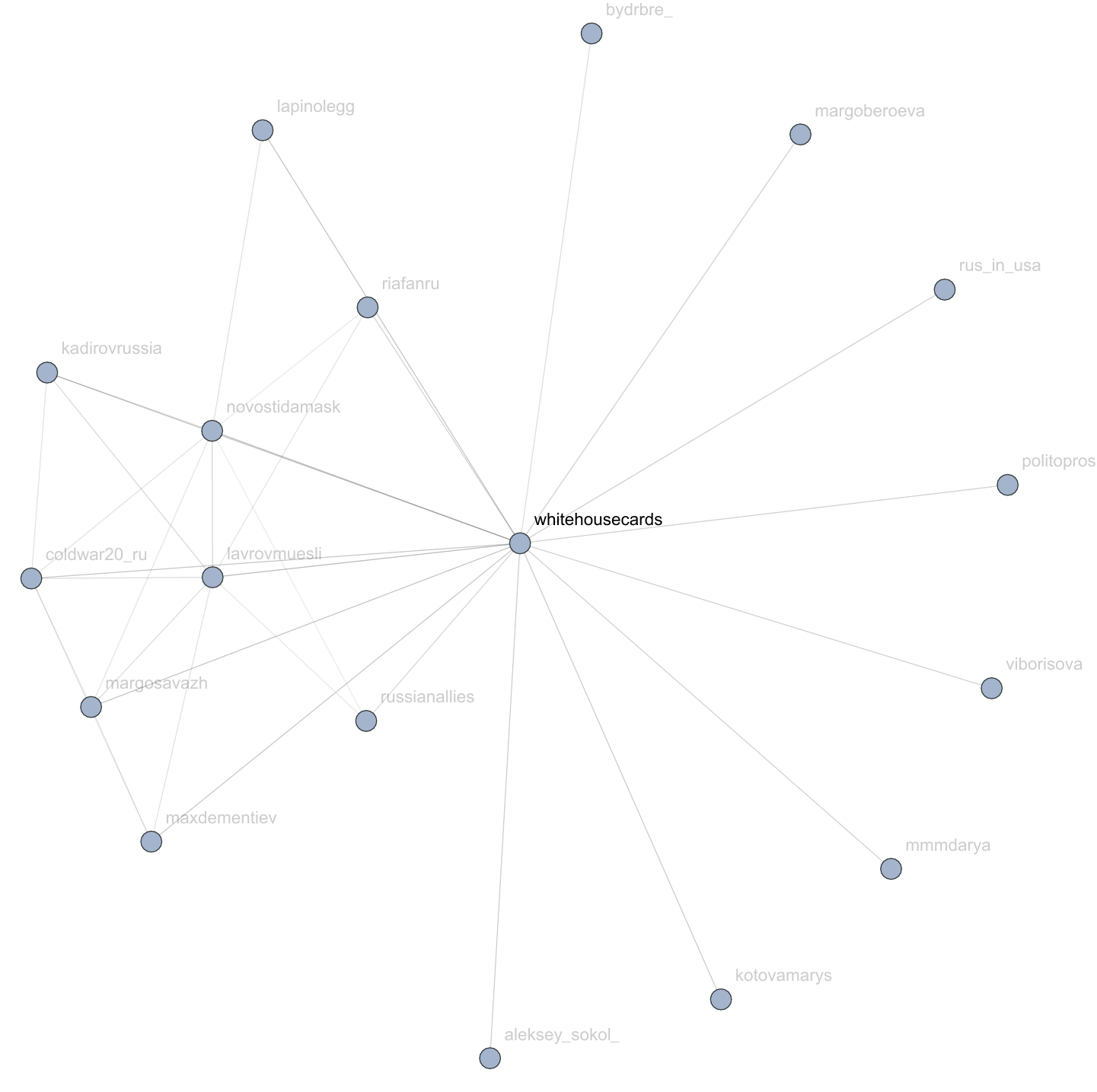}}
  \caption{The user graphs corresponding to both the German language cluster and the Russian language cluster. Notice there are several central users surrounded by other users who follow their lead (linguistically).}
  \label{fig:Clusters}
\end{figure}

In the main text, we refer to the German user cluster generated as a result of language analysis. We illustrate both this cluster and the cluster corresponding to the Russian language user cluster in Figure \ref{fig:Clusters}.

\section*{Word Clouds Broken Out by German Election and French Election}
\begin{figure}[p]
  \centering
  \subfloat[2016 German Election]{\includegraphics[scale=0.5]{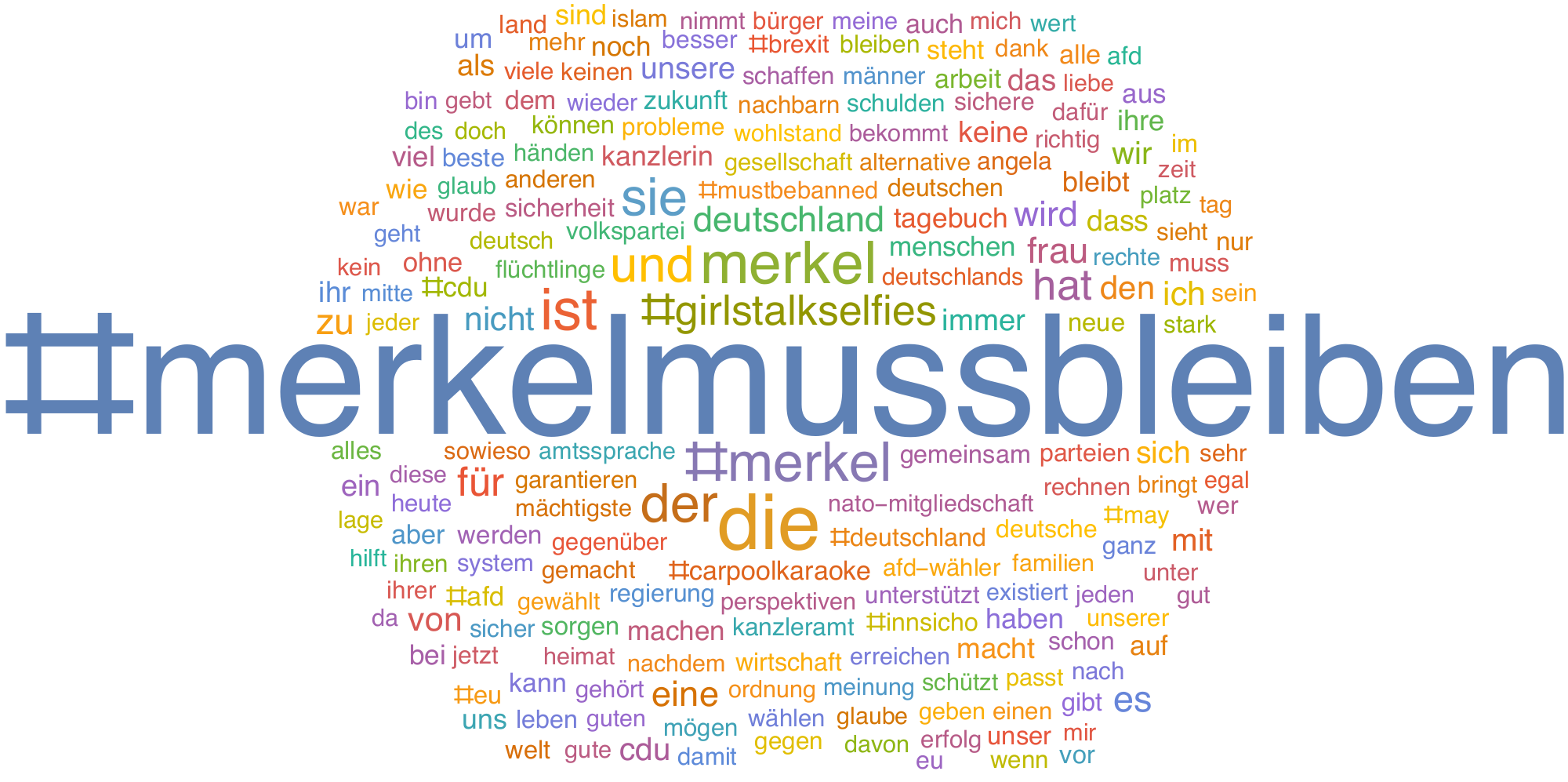}} \qquad
  \subfloat[2017 German Election]{\includegraphics[scale=0.5]{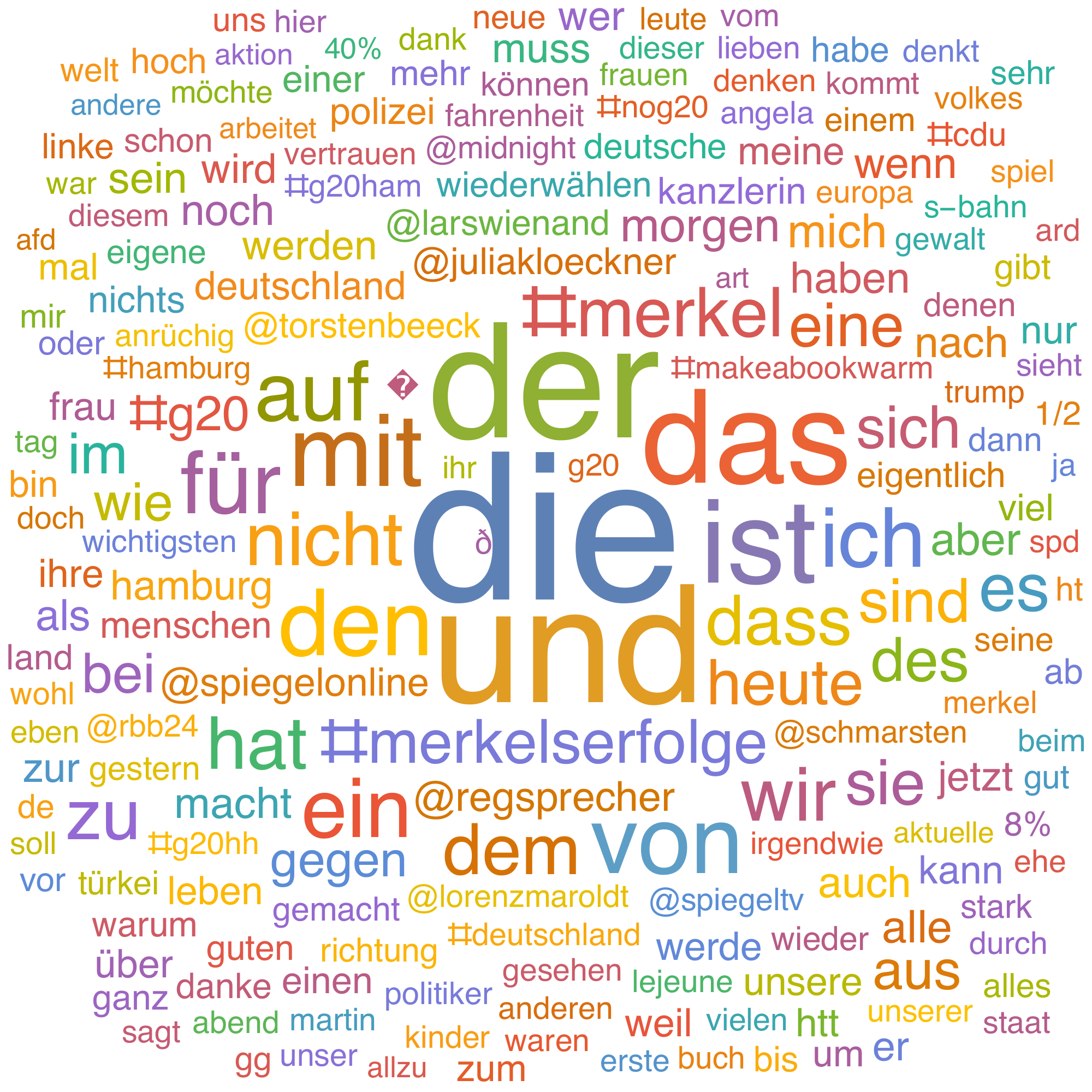}}
  \caption{The two word clouds show distinct characteristics. The 2016 word cloud features a hash tag, apparently supportive of Merkel. The 2017 word cloud is more generic and text normalization may need to be applied in order to identify major topic areas. It is widely believed that \#merkelmussbleiben was generated by bots.}
  \label{fig:GermanWordClouds}
\end{figure}

In the main text, we noted that the German text had distinct characteristics between the 2016 and 2017 elections. Representative word clouds are shown in Figure \ref{fig:GermanWordClouds}.
The picture presented by these word clouds adds to the confusion regarding the nature of the influence campaign (assuming there was an influence campaign) and its goals. The hashtag \#merkelmussbleiben seems to be supportive of a Merkel win in her local Berlin election. (Merkel muss bleiben translates to Merkel must stay.) This contradicts the generally accepted hypothesis that the influence campaigns  were designed to usher in far right-wing governments. On the other hand, \cite{MA16} suggests there is evidence that the users generating this hashtag are bots. It is unclear whether this hashtag is meant ironically or literally. By contrast, the words surrounding the 2017 German election are much more generic. In this case, analysis might be improved by adding German stopwords and proper German stemming. Since the set of tweets in German was relatively small, this is left to future work when a larger data set can be acquired.

\begin{figure}[htbp]
  \centering
  \includegraphics[scale=0.5]{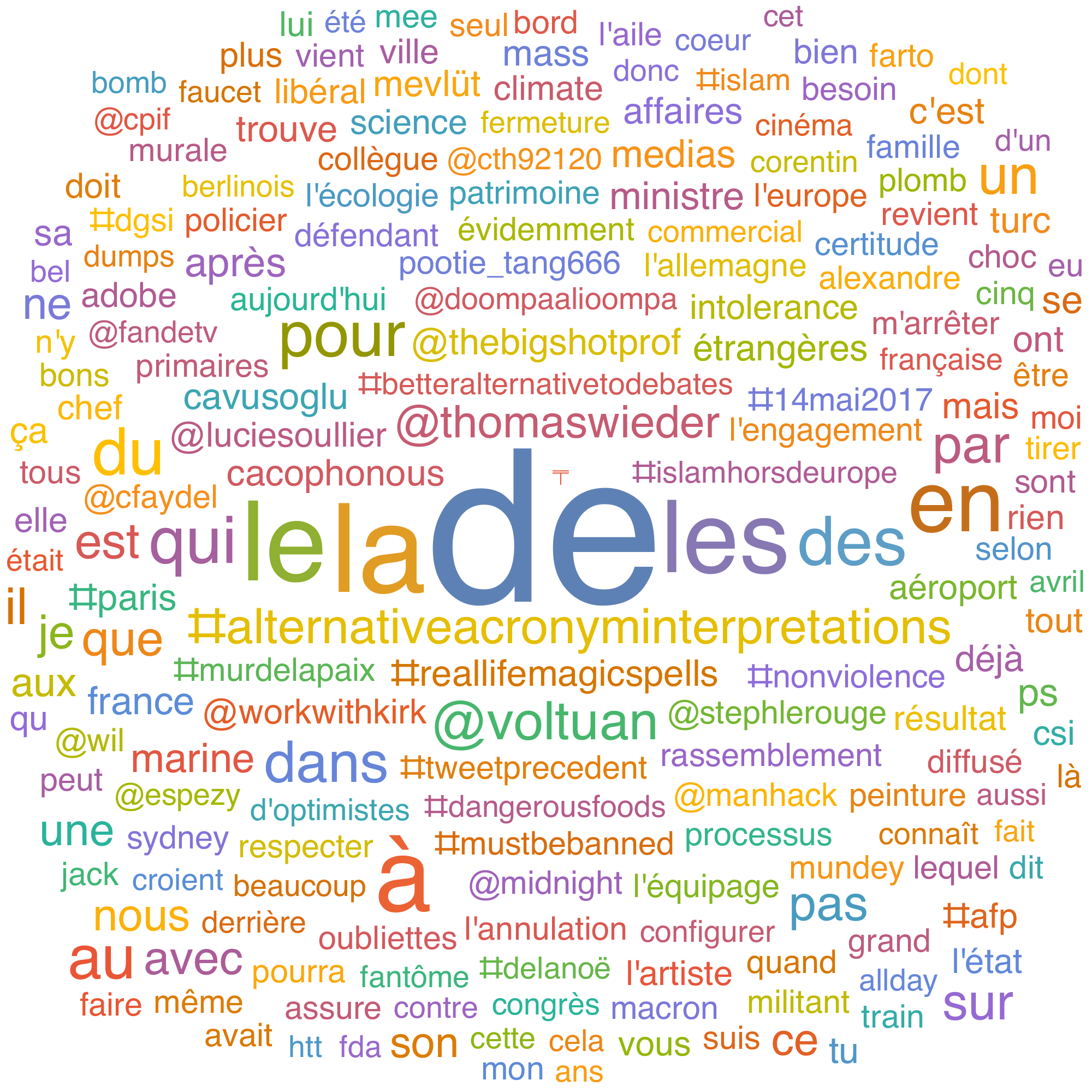}
  \caption{A word cloud of common words posted during the French election in 2017. Clarification by removing stop words may improve the weak signal.}
  \label{fig:FrenchWordClouds}
\end{figure}

A word cloud generated from French tweets posted during the French election is shown in Figure \ref{fig:FrenchWordClouds}.
The French signal is so small that very little can be extracted from the data set. Macron does appear as a topic in the bottom center.

\bibliographystyle{IEEEtran}
\bibliography{RuTwitterReport}

\begin{thebibliography}{10}
\providecommand{\url}[1]{#1}
\csname url@samestyle\endcsname
\providecommand{\newblock}{\relax}
\providecommand{\bibinfo}[2]{#2}
\providecommand{\BIBentrySTDinterwordspacing}{\spaceskip=0pt\relax}
\providecommand{\BIBentryALTinterwordstretchfactor}{4}
\providecommand{\BIBentryALTinterwordspacing}{\spaceskip=\fontdimen2\font plus
\BIBentryALTinterwordstretchfactor\fontdimen3\font minus
  \fontdimen4\font\relax}
\providecommand{\BIBforeignlanguage}[2]{{%
\expandafter\ifx\csname l@#1\endcsname\relax
\typeout{** WARNING: IEEEtran.bst: No hyphenation pattern has been}%
\typeout{** loaded for the language `#1'. Using the pattern for}%
\typeout{** the default language instead.}%
\else
\language=\csname l@#1\endcsname
\fi
#2}}
\providecommand{\BIBdecl}{\relax}
\BIBdecl

\bibitem{J15}
J.~Jasper, ``Social media overtakes entertainment as favourite online
  activity,'' \emph{The Guardian}, vol.
  \url{https://www.theguardian.com/media/2015/sep/02/social-media-overtakes-entertainmemt-as-favourite-online-activity},
  September 1 2015.

\bibitem{P15}
A.~Perrin, ``Social media usage: 2005-2015,'' Pew Research Center, October,
  Available at:
  \url{http://www.pewinternet.org/files/2015/10/PI_2015-10-08_Social-Networking-Usage-2005-2015_FINAL.pdf},
  2015.

\bibitem{BWJT17}
\BIBentryALTinterwordspacing
W.~J. Brady, J.~A. Wills, J.~T. Jost, J.~A. Tucker, and J.~J. Van~Bavel,
  ``Emotion shapes the diffusion of moralized content in social networks,''
  \emph{Proceedings of the National Academy of Sciences}, vol. 114, no.~28, pp.
  7313--7318, 2017. [Online]. Available:
  \url{http://www.pnas.org/content/114/28/7313}
\BIBentrySTDinterwordspacing

\bibitem{RR12}
\BIBentryALTinterwordspacing
J.~M. Rao and D.~H. Reiley, ``The economics of spam,'' \emph{Journal of
  Economic Perspectives}, vol.~26, no.~3, pp. 87--110, September 2012.
  [Online]. Available:
  \url{http://www.aeaweb.org/articles?id=10.1257/jep.26.3.87}
\BIBentrySTDinterwordspacing

\bibitem{P18}
B.~Popken, ``Twitter deleted 200,000 russian troll tweets. read them here.''
  \emph{NBC News}, vol.
  \url{https://www.nbcnews.com/tech/social-media/now-available-more-200-000-deleted-russian-troll-tweets-n844731},
  February 2018.

\bibitem{M18}
T.~Mak, ``{Russia's Divisive Twitter Campaign Took A Rare Consistent Stance:
  Pro-Gun},''
  \url{https://www.npr.org/2018/09/21/648803459/russias-2016-twitter-campaign-was-strongly-pro-gun-with-echoes-of-the-nra},
  September 21 2018.

\bibitem{K18}
A.~Kessler, ``{Who is \texttt{@TEN\_GOP} from the Russia indictment? Here's
  what we found reading 2,000 of its tweets},''
  \url{https://www.cnn.com/2018/02/16/politics/who-is-ten-gop/index.html}, Feb.
  17 2018.

\bibitem{O16}
P.~Oltermann, ``Angela merkel's party beaten by rightwing populists in german
  elections,'' \emph{The Guardian}, September 2016.

\bibitem{TSL00}
\BIBentryALTinterwordspacing
J.~B. Tenenbaum, V.~d. Silva, and J.~C. Langford, ``A global geometric
  framework for nonlinear dimensionality reduction,'' \emph{Science}, vol. 290,
  no. 5500, pp. 2319--2323, 2000. [Online]. Available:
  \url{http://science.sciencemag.org/content/290/5500/2319}
\BIBentrySTDinterwordspacing

\bibitem{BG97}
I.~Borg and P.~Groenen, \emph{Modern Multidimensional Scaling: theory and
  applications,}.\hskip 1em plus 0.5em minus 0.4em\relax New York, NY:
  Springer-Verlag, 1997.

\bibitem{BN02}
M.~Belkin and P.~Niyogi, ``Using manifold structure for partially labeled
  classification,'' in \emph{Advances in neural information processing
  systems}, 2002, pp. 929--936.

\bibitem{FGNP+03}
D.~Friedlander, C.~Griffin, N.~Jacobson, S.~Phoha, and R.~R. Brooks, ``Dynamic
  agent classification using an \textit{ad hoc} mobile acoustic sensor
  network,'' \emph{EURASIP J. Applied Signal Processing}, vol.~4, pp. 371--377,
  March 2003.

\bibitem{LSD12}
\BIBentryALTinterwordspacing
C.~Li, A.~Sun, and A.~Datta, ``Twevent: segment-based event detection from
  tweets,'' in \emph{Proceedings of the 21st ACM international conference on
  Information and knowledge management}, ser. CIKM '12.\hskip 1em plus 0.5em
  minus 0.4em\relax New York, NY, USA: ACM, 2012, pp. 155--164. [Online].
  Available: \url{http://doi.acm.org/10.1145/2396761.2396785}
\BIBentrySTDinterwordspacing

\bibitem{CHBG10}
M.~Cha, H.~Haddadi, F.~Benevenuto, and K.~P. Gummadi, ``Measuring user
  influence in twitter: The million follower fallacy,'' in \emph{in ICWSM 2010:
  Proceedings of international AAAI Conference on Weblogs and Social}, 2010.

\bibitem{WHMW11}
\BIBentryALTinterwordspacing
S.~Wu, J.~M. Hofman, W.~A. Mason, and D.~J. Watts, ``Who says what to whom on
  twitter,'' in \emph{Proceedings of the 20th international conference on World
  wide web}, ser. WWW '11.\hskip 1em plus 0.5em minus 0.4em\relax New York, NY,
  USA: ACM, 2011, pp. 705--714. [Online]. Available:
  \url{http://doi.acm.org/10.1145/1963405.1963504}
\BIBentrySTDinterwordspacing

\bibitem{ZJHS11}
\BIBentryALTinterwordspacing
W.~X. Zhao, J.~Jiang, J.~He, Y.~Song, P.~Achananuparp, E.-P. Lim, and X.~Li,
  ``Topical keyphrase extraction from twitter,'' in \emph{Proceedings of the
  49th Annual Meeting of the Association for Computational Linguistics: Human
  Language Technologies - Volume 1}, ser. HLT '11.\hskip 1em plus 0.5em minus
  0.4em\relax Stroudsburg, PA, USA: Association for Computational Linguistics,
  2011, pp. 379--388. [Online]. Available:
  \url{http://dl.acm.org/citation.cfm?id=2002472.2002521}
\BIBentrySTDinterwordspacing

\bibitem{KJKP13}
\BIBentryALTinterwordspacing
S.~Kim, S.~Jeon, J.~Kim, Y.~Park, and H.~Yu, ``Finding core topics: Topic
  extraction with clustering on tweet,'' in \emph{2012 International Conference
  on Cloud and Green Computing (CGC)}, vol.~00, Nov. 2013, pp. 777--782.
  [Online]. Available: \url{doi.ieeecomputersociety.org/10.1109/CGC.2012.120}
\BIBentrySTDinterwordspacing

\bibitem{SSG13}
A.~Squicciarini, C.~Griffin, and S.~Styer, ``Identifying multi-regime behaviors
  of memes in twitter data,'' in \emph{Proc. Science and Information
  Conference}, London, UK, August 27-29 2014.

\bibitem{N06}
\BIBentryALTinterwordspacing
M.~E.~J. Newman, ``Modularity and community structure in networks,''
  \emph{Proceedings of the National Academy of Sciences}, vol. 103, no.~23, pp.
  8577--8582, 2006. [Online]. Available:
  \url{http://www.pnas.org/content/103/23/8577}
\BIBentrySTDinterwordspacing

\bibitem{LF09}
\BIBentryALTinterwordspacing
A.~Lancichinetti and S.~Fortunato, ``Community detection algorithms: A
  comparative analysis,'' \emph{Phys. Rev. E}, vol.~80, p. 056117, Nov 2009.
  [Online]. Available:
  \url{https://link.aps.org/doi/10.1103/PhysRevE.80.056117}
\BIBentrySTDinterwordspacing

\bibitem{MS99}
C.~D. Manning and H.~Sch\"{u}tze, \emph{{Foundations of Statistical Natural
  Language Processing}}.\hskip 1em plus 0.5em minus 0.4em\relax MIT Press,
  1999.

\bibitem{N16}
\BIBentryALTinterwordspacing
M.~E.~J. Newman, ``Equivalence between modularity optimization and maximum
  likelihood methods for community detection,'' \emph{Phys. Rev. E}, vol.~94,
  p. 052315, Nov 2016. [Online]. Available:
  \url{https://link.aps.org/doi/10.1103/PhysRevE.94.052315}
\BIBentrySTDinterwordspacing

\bibitem{N13}
\BIBentryALTinterwordspacing
------, ``Spectral methods for community detection and graph partitioning,''
  \emph{Phys. Rev. E}, vol.~88, p. 042822, Oct 2013. [Online]. Available:
  \url{https://link.aps.org/doi/10.1103/PhysRevE.88.042822}
\BIBentrySTDinterwordspacing

\bibitem{Z32}
G.~K. Zipf, \emph{{Selected Studies of the Principle of Relative Frequency in
  Language}}.\hskip 1em plus 0.5em minus 0.4em\relax Harvard University Press,
  1932.

\bibitem{M65}
B.~Mandelbrot, ``Information theory and psycholinguistics,'' in
  \emph{Language}.\hskip 1em plus 0.5em minus 0.4em\relax Penguin Books, 1965.

\bibitem{AB00}
\BIBentryALTinterwordspacing
R.~Albert and A.-L. Barab\'asi, ``Topology of evolving networks: Local events
  and universality,'' \emph{Phys. Rev. Lett.}, vol.~85, pp. 5234--5237, Dec
  2000. [Online]. Available:
  \url{https://link.aps.org/doi/10.1103/PhysRevLett.85.5234}
\BIBentrySTDinterwordspacing

\bibitem{MA16}
N.~Meves and K.~Allgemein, ``{Der Bot Boost \#merkelmussbleiben},''
  {https://www.wahl.de/aktuell/2016/08/05/social-bot-bundesregierung/}, May
  2016.

\end{thebibliography}

\end{document}